\documentclass[10pt,aps,onecolumn,superscriptaddress,floatfix,nofootinbib,longbibliography,notitlepage]{revtex4-2}
\usepackage{tikz}
\usepackage{lipsum}
\usepackage{graphicx}
\usepackage[export]{adjustbox}
\usepackage{amsmath,amssymb,wasysym}
\usepackage{braket}
\usepackage{mathtools}
\usepackage{hyperref}
\usepackage{mathrsfs}
\usepackage{verbatim}
\usepackage{cases}

\newcommand{\be}{\begin{align}}
\newcommand{\ee}{\end{align}}

\newcommand{\R}{\mathbb{R}}

\usepackage{color}
\definecolor{Blue}{rgb}{0.00, 0.00, 1.00}
\definecolor{Red}{rgb}{1.00, 0.00, 0.00}
\definecolor{Green}{rgb}{0.00, 0.70, 0.00}
\definecolor{Violet}{rgb}{1.00, 0.0, 1.00}

\newcommand{\blue}{}

\newcommand{\beq}{\begin{equation}}
\newcommand{\eeq}{\end{equation}}
\newcommand{\bea}{\begin{eqnarray}}
\newcommand{\eea}{\end{eqnarray}}

\begin{document}

\title{Clusters in the critical branching Brownian motion}
\author{Beno\^it Fert\'e}
\affiliation{Laboratoire de Physique de l'\'Ecole normale sup\'erieure, ENS, Universit\'e PSL, CNRS, Sorbonne Universit\'e, Universit\'e Paris Cit\'e, F-75005 Paris, France}
 \affiliation{Universit\'e Paris-Saclay, CNRS, LPTMS, 91405, Orsay, France}

\author{Pierre Le Doussal}
	 \affiliation{Laboratoire de Physique de l'\'Ecole normale sup\'erieure, ENS, Universit\'e PSL, CNRS, Sorbonne Universit\'e, Universit\'e Paris Cit\'e, F-75005 Paris, France}

\author{Alberto Rosso}
 \affiliation{Universit\'e Paris-Saclay, CNRS, LPTMS, 91405, Orsay, France}
 \author{Xiangyu Cao}
\affiliation{Laboratoire de Physique de l'\'Ecole normale sup\'erieure, ENS, Universit\'e PSL, CNRS, Sorbonne Universit\'e, Universit\'e Paris Cit\'e, F-75005 Paris, France}
 \begin{abstract}

Brownian particles that are replicated and annihilated at equal rate have strongly correlated positions, forming a few compact clusters separated by large gaps. We characterize the distribution of the particles at a given time, using a definition of clusters in terms a coarse-graining length recently introduced by some of us.
We show that, in a non-extinct realization, the average number of clusters grows as $\sim t^{D_{\mathrm{f}}/2}$ where $D_{\mathrm{f}} \approx 0.22$ is the Hausdorff dimension of the boundary of the super-Brownian motion, found by Mueller, Mytnik, and Perkins. We also compute the distribution of gaps between consecutive particles. We find two regimes separated by the characteristic length scale $\ell = \sqrt{D/\beta}$ where $D$ is the diffusion constant and $\beta$ the branching rate. The average number of gaps greater than $g$ decays as $\sim g^{D_{\mathrm{f}}-2}$  for $g\ll \ell$ and $\sim g^{-D_{\mathrm{f}}}$ for $g \gg \ell$.
Finally, conditioned on the number of particles $n$, the above distributions are valid for $g \ll \sqrt{n}$; the average number of gaps greater than $g \gg \sqrt{n}$ is much less than one, and decays as $\simeq 4 (g/\sqrt{n})^{-2}$, in agreement with the universal gap distribution predicted by Ramola, Majumdar, and Schehr. Our results interpolate between a dense super-Brownian motion regime and a large-gap regime, unifying two previously independent approaches.

 \end{abstract}
\maketitle

\section{Introduction}
The critical branching Brownian motion (BBM) is a simple diffusion-reaction model: non-interacting Brownian particles have the same diffusive constant $D$, and can replicate themselves and be annihilated with an equal rate $\beta$; initially, there is one single particle at the origin.  It is {the} critical point separating two dynamical phases: the sub-critical phase where the annihilation rate is greater and the process rapidly goes extinct; and the super-critical phase with a greater branching rate and an exponentially growing {number of particles}. At criticality, the number of particles is conserved {on} average, but displays strong statistical fluctuations that grow with time. The process survives {up to time $t$} with probability $\propto 1/(\beta t)$; yet, if it survives, the \textit{typical} number of particles is large $n_{\text{typ}} \propto \beta t$. 
\begin{figure}
    \centering
    \includegraphics[scale=.6]{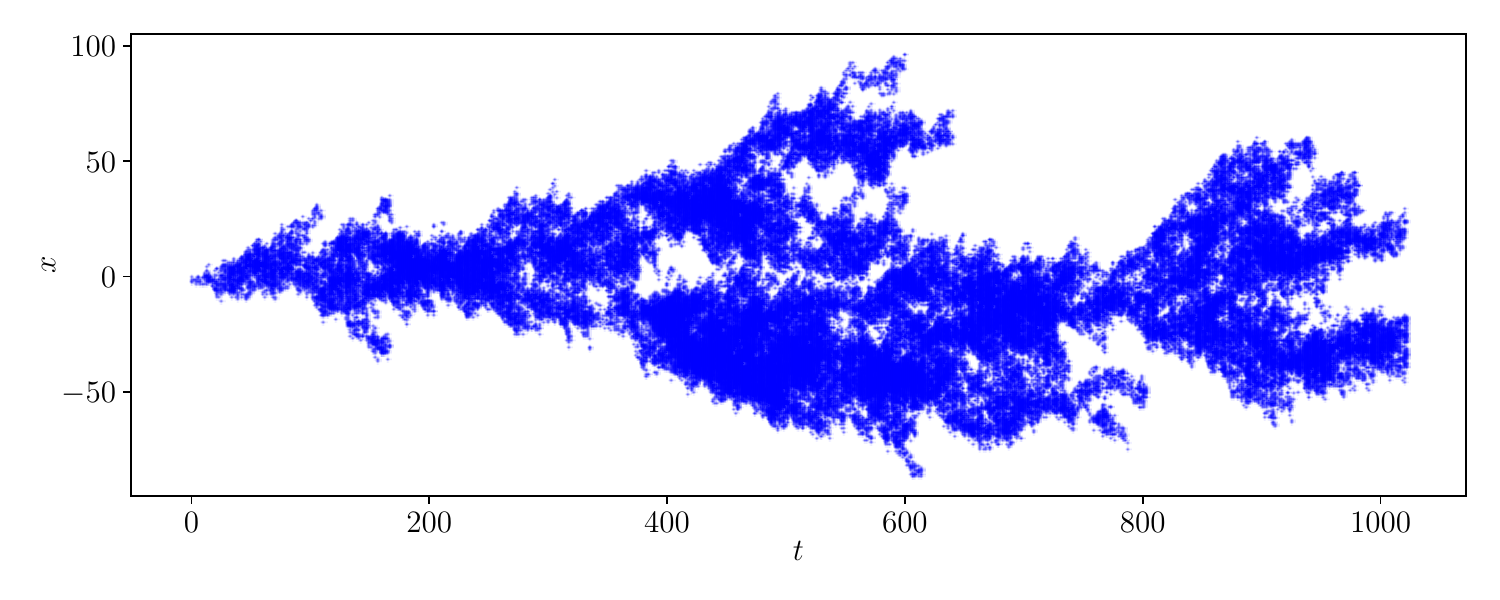}
    \caption{A realization of a discrete analogue of the critical branching Brownian motion (simulated up to $1024$ time steps). In each time step, each particle replicates itself with probability $1/2$,  and is removed otherwise. Then, each particle displaces by an independent standard Gaussian variable. The particles form dense clusters that can be separated by large gaps.}
    \label{fig:sample}
\end{figure}

The positions of these particles are also strongly correlated at criticality. As particle creation takes place only where they already exist, they tend to agglomerate and form clusters. This clustering property (sometimes referred to as patchiness) makes the critical BBM model an interesting model in various domains of physics and biology, such as population dynamics~\cite{zhang90,tsimring,Houchmandzadeh0,Houchmandzadeh,Houchmandzadeh2}, epidemics~\cite{bailey1975mathematical,kendall1956deterministic,dumonteil13extent}, genetics~\cite{meyer96,lawson07}, and neutron scattering in nuclear reactors~\cite{pazsit2007neutron,deMulatier2014,deMulatier2015}.
While its applications often concern higher dimensions, clustering is the strongest in one dimension. Figure~\ref{fig:sample} shows a realization of the 1D BBM in spacetime. The herd of particles form a few dense clusters. They ebb and flow collectively, and are separated by large gaps from time to time. As a first step {to quantitatively understand} the clustering dynamics, we may seek to characterize the spatial distribution of the particles \textit{at a fixed time}. One would like to define and count the clusters, and characterize the size distribution of the gaps between them. These are nontrivial questions, and have been recently pursued by two approaches. 

The first one \cite{ramola,ramola15} focuses on a natural observable: the gaps between consecutive particles. More precisely, the authors considered realizations with a fixed number $n$ of particles {alive} at time $t$, and computed exactly the probability distribution of the gap between the $k$-th and $(k+1)$-th consecutive particles, $k = 1, \dots, n-1$, using a mapping to a hierarchy of Kolmogorov–Petrovsky–{Piskunov}~\cite{kpp} (KPP) equations. They found that this distribution has a well-defined limit as $t\to\infty$. Moreover, the limit probability density function $P_{n, k}(g)$ has a universal tail,
\begin{equation}
     P_{n,k}(g) \simeq \frac{8D}{\beta g^3} \,,\, g\to\infty \,, \label{eq:gap3_intro}
\end{equation}
where the prefactor is exact and the same for any $n$ and $k$. Here and below, $A \simeq B$ means that $A / B$ tends to one in the regime specified; $A \sim B$ means that $A/B$ tends to an unknown constant. 

The second approach, mainly pursued in the probability theory literature, considers a continuum limit of the critical BBM, called the super-Brownian motion, {where the branching rate is so large that 
it is more suitable to consider the system as a random fluid~\footnote{Usually, one also modifies the initial condition and consider a large number of particles at $x= 0$ instead of one. However, if we condition on non-extinction, these two initial conditions are equivalent. See Appendix~\ref{sec:SBM}.}. 
It is then described by a time dependent random density profile $\rho\equiv \rho(x,t)$, which can be quite irregular,} and whose time evolution is governed by a stochastic differential equation~\cite{perkins1,slade2002scaling}. {It was pointed out recently~\cite{pierrenew} that this 
continuum limit of the BBM is equivalent to the Brownian force model, a mean-field theory of the avalanches which occur near the depinning transition of an elastic interface in a disordered medium~\cite{pldw12,pldw13,pierrenewBFM};} in that context, the {local} density corresponds to the {local} velocity of the interface.
The notion of clusters is well-defined in this continuum limit (see below). Indeed, $\rho$ is positive on a number of intervals, separated by gaps where $\rho = 0$. We can view these intervals as the clusters of the SBM. We can count them by measuring the size of their boundary (since every intervals has two boundary points). {It was shown 
in~\cite{mytnik17} (see also \cite{hughes}) that {in this continuum limit} the boundary has a nontrivial fractal dimension, whose value $D_{\mathrm{f}}$ is determined in terms of the leading eigenvalue of a particular Ornstein–Uhlenbeck generator with a killing term. Bounds were provided in
~\cite{mytnik17} and here we determine $D_{\mathrm{f}} \approx 0.22$ by solving numerically
this Ornstein–Uhlenbeck problem (see \eqref{eq:H} below).}
{ We can interpret this result as follows. {\blue As we just explained, the average number of clusters (in a non-extinct realization) is proportional to the size of the boundary of $\mathrm{supp}(\rho) = \{x: \rho(x) > 0\}$. The latter can be in turn estimated as the extent spanned by the diffusing particles, $\propto \sqrt{t}$, raised to the power $D_{\mathrm{f}}$ (by the definition of the fractal dimension):
\begin{equation}
      \left< N_c \right>_{\text{non-ex}} \sim  \left(\sqrt{t}\right)^{D_{\mathrm{f}}} \sim  \, t^{0.11} \,.  \label{eq:Nc_intro}
\end{equation} }
Here, $\left< \dots \right>_{\text{non-ex}}$ denotes an average on non-extinct realizations, and $c$ depends on a short-distance cutoff necessary to make $N_c$ finite [see \eqref{eq:NcbSBM} below]. So, the number of clusters grows as a power law in time, albeit very slowly.
This result {raises} a number of {outstanding} questions concerning the critical BBM model \textit{away from} the continuum limit. How can we define the notion {\blue of clusters from} a finite number of points? If so, can we recover \eqref{eq:Nc_intro}? Moreover, does the exponent $D_{\mathrm{f}}$ inform us about the gaps? Last but not least, the results of the two approaches \eqref{eq:Nc_intro} and \eqref{eq:gap3_intro} seem completely unrelated. How can we reconcile them in a  
{more complete} picture?

\begin{figure}
    \centering
    \includegraphics[scale=.66]{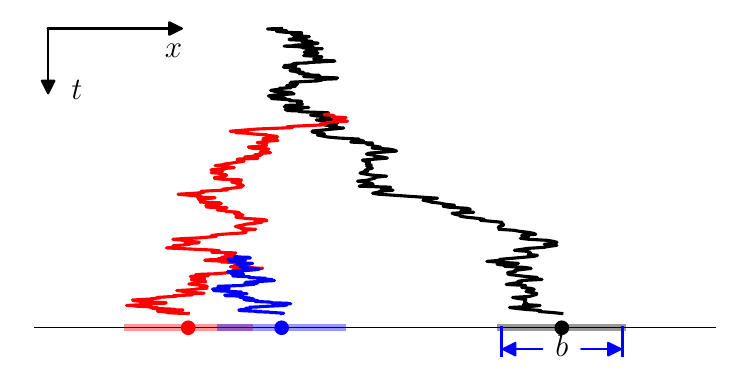}
    \caption{The definition of clusters, illustrated with a realization with $3$ particles (colored for visibility). We surround each of them with an interval of size $b$.  For the value of $b$ illustrated, there are two clusters (red + blue, and black), {i.e. $N_c(b)=2$}. }
    \label{fig:def}
\end{figure}

\section{Summary of results}
\subsection{Method}
In this work, we address these questions using a simple approach recently introduced by some of us~\cite{cao22}. The basic idea is to consider a coarse-grain scale $b$, and assign to each particle an interval of size $b$ centered around it. If two particles {are separated by} a distance less than $b$, their respective intervals will overlap and form {clusters}. See Figure~\ref{fig:def} for an illustration. We shall take this as a definition of clusters and denote the number of clusters by $N_c(b)$, to emphasize its $b$ dependence. It is a decreasing function of $b$. Indeed, for a fixed realization of $n$ particles, $N_c(b \to 0) = n$ as the intervals are too small to overlap with each other. As $b$ increases, and exceeds the size of a gap, two neighboring clusters will merge and $N_c(b)$ will decrease by one. Finally, when $b$ is greater than all the gaps, $N_c(b) = 1$, as all the intervals are connected. In other words, $N_c(b)$ is one plus the number of gaps larger than $b$ for non-extinct realizations ($N_c(b) = 0$ for extinct ones):
\begin{equation}\label{eq:Ncbgaps}
    N_c(b) = \begin{dcases} 
      \text{number of gaps $\ge b$} & n > 0 \text{ (non-extinct)}\\
      0 & n = 0  \text{ (extinct)}
    \end{dcases}
\end{equation}
Therefore its dependence on $b$ is closely related to the distribution of the gaps. More precisely, in terms of $P_{n,k}(g)$, the probability density function of the $k$-th gap conditioned on there being $n$ particles, we have
\begin{equation} \label{eq:NcbPg}
  \left< N_c(b) \right>_n = 1+ \sum_{k=1}^{n-1} \int_b^\infty P_{n,k}(g) \mathrm{d} g \,,
\end{equation}
where $ \left< [...]\right>_n $ denotes an average conditioned on the number of particles at time $t$.

\subsection{Typical realizations}\label{sec:res_typ}
A first result of this work is to understand the large time $\beta t \gg 1$ behavior of $N_c(b)$ of a typical non-extinct realization. Here, we heuristically derive it with a scaling argument combined with \eqref{eq:Nc_intro} (a more refined analysis presented later will confirm it).  To do this, we identify the relevant length scales of the problem. They are the following, in decreasing order: 
\begin{enumerate}
    \item the extent spanned by all the particles $\xi \sim \sqrt{D t}$, i.e., the typical displacement of the leftmost and rightmost particle of the critical BBM~\cite{ramola}.
    \item the ``mean free path'' of a particle between consecutive branching/annihilation events, $\ell = \sqrt{D / \beta}$. This length scale becomes $1$ in the natural units $D = \beta = 1$. 
    \item the inverse density, $a \sim  \xi / n_{\text{typ}} \sim \sqrt{D / (\beta^2 t)} $. 
\end{enumerate}
By dimensional analysis, $N_c(b)$ must be a function of the two dimensionless quantities of the problem, $\beta t$ and $b / \ell$. Intuitively, we expect $N_c(b) \sim n_{\text{typ}} \sim \beta t$ for very small gaps $b \lesssim a$, $N_c(b) \sim 1$ for very large gaps $b\gtrsim \xi$; Thus $N_c(b)$ can have the nontrivial $t$ dependence \eqref{eq:Nc_intro} only at the remaining relevant length scale $b\sim \ell$: $N_c(b\sim \ell) \sim  (\beta t)^{D_{\mathrm{f}} / 2}$; the $\beta$ dependence here is fixed by the dimensional analysis. Finally, it is natural to expect that $N_c(b)$ behaves as a power law in each of the scaling regimes $a \ll b \ll \ell$ and $\ell \ll b \ll \xi$. 
Gathering the above information, we surmise that:
\begin{equation}
    \left< N_c(b)\right>_{\text{non-ex}} \sim \begin{dcases}
    (b / \ell)^{D_{\mathrm{f}} - 2} (\beta t)^{\frac{D_{\mathrm{f}}}2} =  \beta^{D_{\mathrm{f}}-1} D^{1-\frac{D_{\mathrm{f}}}2} \, t^{\frac{D_{\mathrm{f}}}2} \, b^{D_{\mathrm{f}} - 2}  &
        a  \lesssim b \lesssim \ell \\
         (b / \ell)^{-D_{\mathrm{f}}} (\beta t)^{\frac{D_{\mathrm{f}}}2} = D^{\frac{D_{\mathrm{f}}}2} \, t^{\frac{D_{\mathrm{f}}}2} \, b^{-D_{\mathrm{f}}} 
         & \ell \lesssim b \lesssim  \xi 
    \end{dcases}  \,. \label{eq:Ncb_intro}
\end{equation}
In other words, most gaps of a typical realization fall into two categories, each governed by a different power law distribution. To understand this intuitively, we may inspect Figure~\ref{fig:sample} once more in light of the formula \eqref{eq:Ncb_intro}. Many small gaps are not visible in that Figure. 
Eq.~\eqref{eq:Ncb_intro} predicts that their sizes are governed by a rather large exponent $N_c(b) \propto b^{D_{\mathrm{f}} - 2} \approx b^{-1.78}$. Most such gaps are comparable to the inverse density $\sim 1/\sqrt{t}$. Meanwhile, the large gaps that are visible in Figure~\ref{fig:sample} have a wide distribution, $ N_c(b) \propto b^{D_{\mathrm{f}} - 2} \approx b^{-0.22}$. These predictions are verified numerically, see Figure~\ref{fig:numerics} in Section~\ref{sec:num} below.

We now make connection with the continuum limit of the super-Brownian motion (SBM) and of the Brownian force model. For this, we take $\beta$ to infinity, while keeping $D$ of order one in \eqref{eq:Ncb_intro} so that the typical interparticle distance scale $a \to 0$. Our result \eqref{eq:Ncb_intro} becomes the following:
\begin{equation}
    \left< N_c(b)\right>_{\text{non-ex}} \sim D^{\frac{D_{\mathrm{f}}}2} \, t^{\frac{D_{\mathrm{f}}}2} \, b^{-D_{\mathrm{f}}} \,,\, b \ll \sqrt{D t} \, \text{  (SBM, or $\beta \gg 1$ limit)} \label{eq:NcbSBM}
\end{equation}
Indeed, the small-gap regime of \eqref{eq:Ncb_intro} disappears ($\ell \to 0$),  and the large-gap one extends to all values of $b \lesssim \sqrt{Dt}$ and remains well-defined, being independent of $\beta$. The $t$ dependence of \eqref{eq:NcbSBM} and \eqref{eq:Nc_intro} are the same, yet \eqref{eq:NcbSBM} has the correct dimension. Its $b$-dependence should be interpreted as describing the size distribution of the gaps between neighboring connected components of the support of the density $\rho$. In the SBM limit, there are infinitely many such gaps, which can be arbitrarily small. On the other hand, the average number of gaps $\ge b$ is finite, and decays as $b^{{-D_{\mathrm{f}}} }$. We can view $b$ as the short-distance cutoff that is necessary to have a finite number of clusters. This cutoff is different from that introduced in \cite{mytnik17}, which is a cutoff on the value of the local density. In that sense the above result is new, and it is interesting to note that it also applies, as mentioned above, to the gaps in the support of the velocity in a mean-field avalanche.

\subsection{Particle number conditioning}
The puzzle remains with regard to the first approach~\cite{ramola,ramola15}: eq.~\eqref{eq:Ncb_intro} still looks unrelated to~\eqref{eq:gap3_intro}. A second result of this work reconciles the two approaches by calculating the averaged cluster number, conditioned on a fixed number of particles $n$ (at time $t$). We shall show that, the conditioned average number of clusters has a long time limit with the following asymptotic behaviors for $n \gg 1$:
\begin{equation}\label{eq:Ncbn_intro}
    \left< N_c(b) \right>_n - 1 \stackrel{t \to\infty}\sim \begin{dcases} 
   c (b / \xi_n)^{-D_{\mathrm{f}}}  & 1 \ll b  \ll \xi_n \\
   4 (b / \xi_n)^{-2}  & b \gg \xi_n 
    \end{dcases} \,,\, \xi_n = (D n / \beta)^{\frac12}  \,.
\end{equation}
Here, the crossover scale $\xi_n$ can be understood as the typical extent of a realization with $n$ particles. The result in the $b \ll \xi_n$ regime involves an unknown factor $c$ that is independent of $b$. Yet, the prefactor of the $b \gg \xi_n$ regime is exact. 

Eq.~\eqref{eq:Ncbn_intro} reconciles the result~\eqref{eq:gap3_intro} of \cite{ramola} and \eqref{eq:Ncb_intro} just announced. Indeed, the {formula for} $b \gg \xi_n$ regime is consistent with~\eqref{eq:gap3_intro}. To see why, recall that $ \left< N_c(b) \right>_n - 1 $ is the average number of gaps greater than $b$. Then, Eq.~\eqref{eq:gap3_intro} implies that each of the $n-1$ gaps is larger than $b$ with probability $4 D / (\beta b^2)$. So the average total number of gaps greater than $ b$ is $4 (n-1) D/ (\beta b^2)$ by \eqref{eq:NcbPg}~\footnote{Note that we do not need to assume independence between gaps, since we are calculating a sum of expectation values, not a product.}, which is asymptotically equivalent --- including the prefactor --- to \eqref{eq:Ncbn_intro} as $n \gg 1$. Also, the {formula for} $b \ll \xi_n$ is consistent with our result \eqref{eq:Ncb_intro} without particle number conditioning: by setting $n$ to be the typical {number of particles alive} at $t$, $n_{\rm typ} \sim \beta t$, we recover the $b \gg 1$ regime of \eqref{eq:Ncb_intro}. 

Last but not least, \eqref{eq:Ncbn_intro} provides an additional insight on the distinct nature of the two approaches. It implies in particular $ \left< N_c(b = \xi_n) \right>_n \sim 1$, i.e., in a typical realization of $n \gg 1$ particles, the largest gap size is comparable to $\xi_n$. If we take a \textit{single} typical configuration with  $n\gg1$ particles and make a histogram of its $n-1$ gaps, according to \eqref{eq:Ncbn_intro}, it will display a $P(g) \sim g^{-D_{\mathrm{f}} - 1}$ distribution with a cutoff at $g \sim \xi_n$. Meanwhile, if we consider \textit{many} realizations with $n$ particles, large gaps $g \gg \xi_n$ will appear from time to time. A histogram of the gaps of all the realizations combined will show a $P(g) \sim g^{-3}$ power law for $g \gg \xi_n$. Observing this numerically is very challenging, especially for large $n$. Nevertheless, thanks to the knowledge of the exact prefactor, Ref.~\cite{ramola} was able to identify the $g^{-3}$ regime for $n\le 10$.  \vspace{.4cm}

The rest of the paper is organized as follows. In Section~\ref{sec:kpp} we analyze the model using a mapping to a KPP equation, and derive the predictions announced above. Section~\ref{sec:num} reports a few numerical tests supporting the predictions. We close with concluding remarks in Section~\ref{sec:conclude}. The main text is supplemented with Appendices containing analytical details and numerical methods. 

\section{Analysis of the KPP equation}\label{sec:kpp}
\subsection{Setting up the mapping}
In this section we set up a mapping to a set of KPP equations that allows to calculate $\left< N_c(b) \right>$ and related quantities in the critical BBM model. From now on we work with the natural units $D = \beta = 1$. This amounts to measure time in the unit of the inverse branching rate $1/\beta$ and length in the unit of the mean free path $\sqrt{D/\beta}$. Thus, the dependence on $D$ and $\beta$ can be readily restored. 

 Let $x_1(t), \dots, x_n(t)$ be the position of the particles; here $n = n(t)$ is the {\blue number of particles}, and is a random variable. Consider the following observable  (here and below, $\left< [\dots] \right>$ denotes an average on all realizations, including those already extinct by time $t$; we recall that the initial condition is that of a single particle at the origin) : 
\begin{equation}
  E(x,t) := \left< \prod_{i=1}^{n} f(x - x_i(t)) \right> \,,
\end{equation}
$f$ is any function (if $n = 0$, the product is equal to $1$ by convention). A standard backward recursion argument (see Appendix \ref{app:kpp}) shows that  $E(x,t)$ satisfies the following equation 
\begin{align}
    \partial_t E = \partial_{x}^2 E + (1-E)^2 \label{eq:kpp} \,,\, 
    E\vert_{t = 0} = f(x) \,.
\end{align}
 It is often convenient to rewrite the KPP equation by considering
\begin{equation}
    F(x,t) := 1 - E(x,t) \,,
\end{equation}
in terms of which \eqref{eq:kpp} becomes particularly simple at criticality:
\begin{equation}\label{eq:KPPF_sec2}
    \partial_t F = \partial_x^2 F - F^2 \,,\, F\vert_{t=0} = 1 - f(x) \,.
\end{equation}

The above discussion applies to any $f$. As a warm-up example, let us consider $f(x) = e^{-\mu}$, which is independent of $x$. Then $F(x,t) = 1 - \left< e^{-\mu n} \right>$ is a moment generating function of the particle number $n = n(t)$. The KPP equation reduces to an ODE $\dot{F} = - F^2, F(0) = 1 - e^{-\mu}$. the solution is 
\begin{equation}\label{eq:F0}
    F(x, t) = \frac{1}{(1-e^{-\mu})^{-1} + t} := F_0 
\end{equation}
(We gave the expression a name since it will appear a few times.)  One can check that this corresponds to the following well-known distribution of particle number:
\begin{equation} \label{eq:Pn}
    \mathbb{P}(n(t) = 0) = 1 - \frac{1}{t+1} \,,\,   \mathbb{P}(n(t) = k > 0) = \frac{ t^{k-1}}{(1 + t)^{k+1}} \stackrel{t \gg 1}\sim \frac1{t^2} e^{- k / t} \,.
\end{equation}
 
In this work, we shall focus on the following choice of $f$ that allows to make connection with the quantity $N_c(b)$:
\begin{equation}\label{eq:f}
    f(x) = e^{-\mu} \theta(|x| - b/2) 
\end{equation}
so that 
\begin{equation} \label{eq:Ext}
    F(x,t) = 1 - \left< \prod_{i=1}^n \left( e^{-\mu} \theta(|x - x_i(t)| - b/2) \right) \right> \,.
\end{equation}
In plain words, a configuration $x_1 < \dots < x_n$ contributes $1$ to $F$ if $x$ belongs to {at least one} of the intervals $[x_i - b/2, x_i + b/2]$; otherwise, {if $x$ is outside the union of these intervals}, the contribution is $1 - e^{-\mu n}$ 
where $n = n(t)$ is the number of particles. As a consequence, integrating $F$ over a large interval $|x| < L$ gives
\begin{equation}
     \int_{-L}^L F(x) \mathrm{d} x = 2 L F_0 + \left<  e^{-\mu n}  \ell(b) \right> \,.
\end{equation}
 where {$F_0=1 - \left< e^{-\mu n} \right>$} is given above \eqref{eq:F0}. We {have} introduced another geometric quantity of interest, $\ell(b)$, called the \textit{extension}, defined as the total length of the union of the intervals (the overlaps {being} counted only once).  It is not hard to see that 
 \begin{equation}\label{eq:Flimit}
     F(|x|\to\infty) \to F_0 =  \frac{1}{(1-e^{-\mu})^{-1} + t} \,.
 \end{equation}
To get rid of diverging part $2 L F_0$, we can {\blue subtract $F$ by $F_0$} and take $L \to +\infty$  :
\begin{equation}\label{eq:ellbF}
    \left< \ell(b) e^{-\mu n} \right> = \int_{-\infty}^{+\infty} (F(x,t) - F_0) \mathrm{d} x \,.
\end{equation}

 How does this relate to the number of clusters? Thanks to a geometric relation: $N_c(b)$ can be obtained by deriving $\ell(b)$ with respect to $b$:
 \begin{equation}
     N_c(b) = \partial_b \ell(b) \,. \label{eq:Ncellc}
 \end{equation} 
 To see why this is so, consider the change of $\ell(b)$ as $b$ increases by an infinitesimal amount $\mathrm{d} b$. The contributions come from the two extremities of each cluster, which expand by $\mathrm{d} b/2$, therefore, $\mathrm{d} \ell(b) = 2 N_c(b)\mathrm{d} b/2 =N_c(b)\mathrm{d} b$. This is exactly \eqref{eq:Ncellc}.  
 
Combining \eqref{eq:Ncellc} and \eqref{eq:ellbF}, we obtain
\begin{equation}\label{eq:NcbF}
  \left< N_c(b) e^{-\mu n}  \right> =  \int_{-\infty}^{+\infty}  \partial_b F(x,t) \mathrm{d} x \,.
\end{equation}
The equations \eqref{eq:ellbF} and \eqref{eq:NcbF} allow us to extract geometric information from the solution of the KPP equation. The presence of $e^{-\mu n}$ will allow us to obtain $\left<  N_c(b)  \right>_n$ the mean cluster number conditioned on the number of particles, by an inverse Laplace transform and using \eqref{eq:Pn}. More precisely, for any observable $\mathcal{O}$, if its average conditioned on particle number $n$ depends on $n$ as a power law, 
\begin{equation} \label{eq:On}
    \left< \mathcal{O} \right>_n \simeq n^a  \,,\, n \gg 1 \,,
\end{equation}
using \eqref{eq:Pn}, we have
\begin{equation} \label{eq:Omu}
    \left< \mathcal{O} e^{- \mu n} \right> \simeq \frac1{t^2} \int_0^\infty e^{-\mu n - n / t} n^a  \mathrm{d} n \simeq \frac{ \Gamma(a+1)}{t^2 \mu^{a + 1}}   \,,\, 1 \ll \mu^{-1} \ll t \,.
\end{equation}
The analysis of the KPP equation will provide a result in the form of \eqref{eq:Omu}, which we can readily translate to \eqref{eq:On}.

\subsection{Overview of the analysis}
In the rest of this section, we will analyze the following KPP equation:
\begin{equation}\label{eq:KPPF}
    \partial_t F = \partial_x^2 F - F^2 \,,\, F\vert_{t=0} = \begin{cases} 
     1 & |x| \le b / 2 \\
     1 - e^{-\mu} & |x| > b / 2 
    \end{cases} 
\end{equation}
The initial condition will be often referred to as a ``plateau'' of width $b$. Our goal is to establish the asymptotic behavior of the solution in a number of regimes, in order to derive the results announced in the introduction. We will proceed in two main steps. First, in section~\ref{sec:mu0}, we will set $\mu = 0$, and consider $1 \ll b \ll \sqrt{t}$ (typical large gap regime) as well as $1 \ll b^{-1} \ll \sqrt{t}$ (typical small gap regime). This analysis will lead to the result \eqref{eq:Ncb_intro}. Technically, it will crucially involve the Onstein-Uhlenbeck generator introduced in \cite{mytnik17}. Next, in section~\ref{sec:mu}, we will set $\mu > 0$, and consider various regimes where $b\gg 1$, $t\gg1$ and $\mu \gg 1$, with the goal of deriving \eqref{eq:Ncbn_intro}. 

\subsection{Typical regimes ($\mu = 0$)}\label{sec:mu0}
In this section, we shall consider the KPP equation \eqref{eq:KPPF} with $ \mu = 0$. Then, the initial condition $F\vert_{t=0}= \theta(b/2-|x|)$ becomes zero outside the plateau: it resembles a Dirac delta function viewed from a scale $\gg b$. Indeed, we find numerically that the solution with a plateau initial condition has a similar long-time behavior as that with a delta initial condition, which has been well studied in the mathematical literature~\cite{Brezis,mytnik17}. We will first review the latter, and then discuss how to adapt the theory to our initial condition.

\subsubsection{Delta initial condition: a review}\label{sec:mu0delta}
Consider the KPP equation with a delta function initial condition, 
\begin{equation} \label{FKPPdelta} 
     \partial_t F = \partial_x^2 F - F^2 \,,\, \lim_{t\to0} F = \lambda \delta(x) \,.
\end{equation}
This equation has no interpretation in BBM  but appears naturally in the SBM limit~\cite{mytnik17} and we will follow here the approach of that work. A main result thereof is that the solution has the following long-time asymptotic Ansatz:
\begin{equation}
    F(x,t) = \frac1t \left[  f(x / \sqrt{t}) + C t^{-\eta} \varphi(x / \sqrt{t}) + \dots \right] \,. \label{eq:Ansatz}
\end{equation}
Let us explain this equation in detail. 

The leading term $t^{-1} f(x / \sqrt{t})$ should be a solution to the KPP equation \eqref{eq:KPPF} itself. This imposes the following equation on $f(y)$:
\begin{equation}
    f'' + \frac{y}2 f' - f^2 + f = 0 \,. \label{eq:ODEf}
\end{equation}
This ODE has a unique nonzero solution that decays to $0$ as $y \to \pm \infty$. It can be also specified by the initial conditions $f'(0) = 0$ and $f(0) = 0.6898\dots$. It resembles a Gaussian, and has a tail {$f(y) \sim |y| \exp(-y^2 / 4)  $} as $|y| \to \infty$ (see Figure~\ref{fig:fandphi} for a plot). 

\begin{figure}
    \centering
    \includegraphics[width=.6\textwidth]{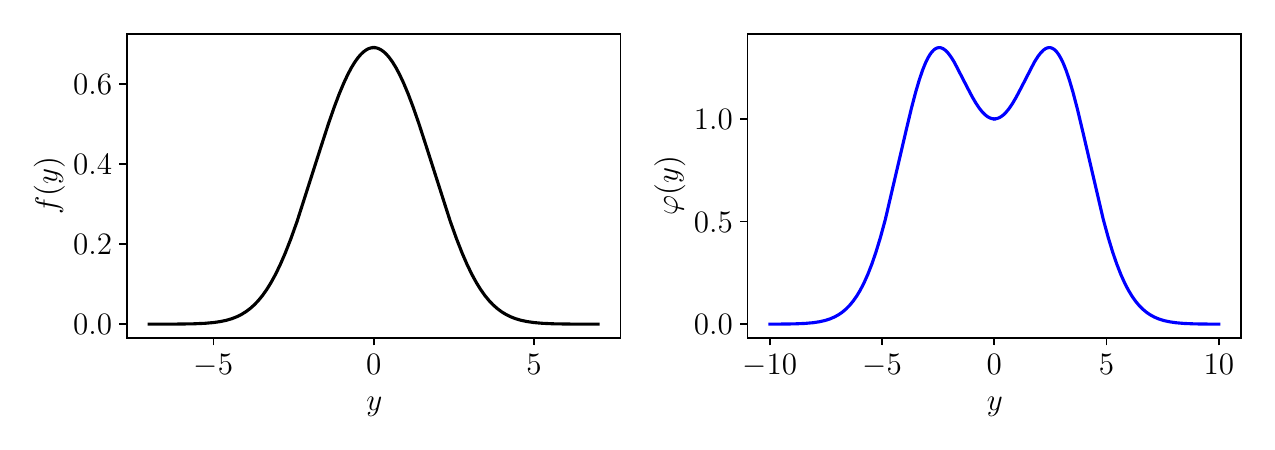}
    \caption{The functions $f$ and $\varphi$ in the Ansatz \eqref{eq:Ansatz}. They are obtained by solving \eqref{eq:ODEf} and \eqref{eq:H}, respectively, using a standard numerical shooting method. The interpretation of these functions in the BBM is discussed below after \eqref{eq:dense1-main}. }
    \label{fig:fandphi}
\end{figure}

The subleading term satisfies a linearisation of the KPP equation around the leading term. One can express this as a an eigenvalue equation involving an operator $H$ {acting on the function $\varphi(y)$}:
\begin{equation}
       \eta \varphi = H \varphi   \,,\,  H = - \partial_y^2 - \frac{y}{2} \partial_y  + 2 f(y) - 1 \,. \label{eq:H}
\end{equation}
Although $H$ is not Hermitian, it is a generator of an Ornstein–Uhlenbeck process with a killing term and can be transformed into a Schr\"odinger Hamiltonian:
\begin{equation} \label{eq:HS}
  {H}_S := e^{y^2 / 8} H e^{-y^2 / 8} = - \partial_y^2 + \frac{y^2}{16} + 2f(y) - \frac{3}4 \,.
\end{equation} 
Therefore, $H$ has all the spectral properties of a standard Schr\"odinger Hamiltonian. Because the potential is confining, it has a discrete spectrum.  We determine $\eta$ and $\varphi(y)$ as the ground state energy and eigen-function, respectively~\footnote{Note that there is another eigenfunction of $H$, with negative energy. It tends to a nonzero constant as $y\to\infty$, and maps to a non-normalizable wave-function of $H_S$. That eigenfunction corresponds to an instability, and will be relevant in the crossover to a $\mu > 0$ solution, see Appendix~\ref{sec:crossover} below. Here, we shall restrict to decaying eigenfunctions of $H$ that correspond to normalizable wavefunctions of $H_S$.}. $\varphi(y)$ has a Gaussian decay $\sim e^{-y^2/4}$ as $|y| \to \infty$ (which is determined by the diffusion term of the KPP equation). We fix the overall normalization of $\varphi$ by requiring that $\varphi(0) = 1$. The value of $\eta$ is not known in a closed form; we estimated it numerically as equal to 
\begin{equation} 
    \eta \approx 0.3904 \dots \,. \label{eq:eta-value}
\end{equation}
 We note that $\eta$ is related to the exponent $\lambda_0$ in \cite{mytnik17}, and to the fractal dimension $D_{\mathrm{f}}$ by the following (\cite{mytnik17}, Theorem 1.3):
\begin{equation}
    \eta = \lambda_0 - 1/2 \,,\, D_{\mathrm{f}} = 2 - 2\lambda_0 = 1-2\eta \approx 0.22 \,, \label{eq:etaDf}
\end{equation}  
as quoted in the introduction. The eigen-function $\varphi(y)$ is plotted in Figure~\ref{fig:fandphi}. Its shape does not resemble a Gaussian, and has a pair of maxima away from $0$.

The linear Schr\"odinger equation \eqref{eq:H} alone cannot fix the prefactor $C$ of the subleading term in \eqref{eq:Ansatz}. Instead, $C$ must be determined by the initial condition. With $\delta$-function initial conditions, this can be done by a scaling argument. Indeed, the rescaling transformation
\begin{equation}\label{eq:rescaling}
     t = \alpha^2 \tilde{t} \,,\, x = \alpha \tilde{x} \,,\, F = \alpha^{-2} \tilde{F} 
     \end{equation}
leaves the KPP equation $\partial_t F = \partial_x^2 F - F^2$ invariant, i.e., one obtains the same equation with tildes. The rescaling transforms the initial condition {in \eqref{FKPPdelta}} 
as follows: $\lambda = \alpha^{-1} \tilde{\lambda} $. Also, the Ansatz \eqref{eq:Ansatz} transforms to itself, which implies $C = \alpha^{2\eta} \tilde{C}$. The above two equations together impose the following scaling law:
\begin{equation}
     F_{t=0} = \lambda \delta(x) \implies 
    C \propto - \lambda^{-2\eta} \label{eq:Clambda}
\end{equation} 
Note that when $\lambda\to\infty$, $C \to 0$. The minus sign of \eqref{eq:Clambda} must be there since $C$ must increase with $\lambda$ (this is because the KPP equation is monotonous: if $F(x,t=0) \ge F'(x,t=0) $ for all $x$, then $F(x,t) \ge F'(x,t)$ for all $t$ and $x$).

\subsubsection{Plateau initial condition}
Having reviewed in some detail the solution to the KPP equation with a delta initial condition, we come back to our plateau initial condition of width $b$.  We expect the solution to obey the Ansatz \eqref{eq:Ansatz} in the long time limit, with however a different prefactor $C$, which now depends on $b$. We now determine that dependence, as well as the time scale above which the Ansatz~\eqref{eq:Ansatz} is valid, for small and large $b$. 

For small gaps, $b \ll 1$, the plateau is indistinguishable from a delta peak with $\lambda = b$. Therefore 
\begin{equation}
      C \sim - b^{-2\eta} \,,\, b \ll 1 \,. \label{eq:Cbsmall}
\end{equation}
We now argue that the asymptotic form~\eqref{eq:Ansatz} is valid only when $t \gg b^{-2}$ for small $b$. Indeed, the initial condition is very singular, so the diffusive term will dominate (against the nonlinear term) initially, leading to a Gaussian solution with $F(0,t) \sim b/\sqrt{t}$. Matching that with the behavior $F(0,t) \sim 1/t$ in \eqref{eq:Ansatz} gives us a crossover time $t \sim b^{-2}$, beyond which \eqref{eq:Ansatz} is valid. 

The case of large gaps, $b \gg 1$, is different. Indeed, the asymptotic Ansatz \eqref{eq:Ansatz} {holds} when and only when $t \gtrsim b^2$; at earlier times, the solution { $F(x,t)$ remains almost independent of $x$ away from the boundaries of the plateau. This changes when $t = O(b^2)$ the time for diffusion over a distance of order the size of the plateau.}. Having identified this time scale, we can determine $C$ by a scaling argument: the relative amplitude of the subleading term, $C t^{-\eta}$, should be of order unity at the only relevant time scale $t\sim b^2$. This fixes 
\begin{equation}
    C \sim b^{2\eta} \,,\, b \gg 1 \,. \label{eq:Cb}
\end{equation}
The sign must be positive here since $C$ must increase with $b$. 

In summary, the long time asymptotic solution to the KPP equation is given by \eqref{eq:Ansatz} with the prefactor $C$ satisfying \eqref{eq:Cbsmall} and \eqref{eq:Cb}:
\begin{equation} \label{eq:Ffinal}
     F(x,t) = \frac1t \left[  f(x / \sqrt{t}) + C t^{-\eta} \varphi(x / \sqrt{t}) + \dots \right] \,,\, C = C(b) \sim  \begin{dcases}
     - b^{-2\eta} &  1 \gg b \gg 1/\sqrt{t} \\
      b^{2\eta} &   1 \ll b \ll \sqrt{t}
     \end{dcases} 
\end{equation}
Applying \eqref{eq:ellbF} {(with $\mu=0$ and $F_0=0$)} to this equation we obtain the average extension
\begin{equation}
    \left< \ell(b)  \right> \sim c_0 t^{-\frac12} + C(b) t^{-\frac12 - \eta} + \dots \,  ,\quad  1/\sqrt{t} \ll b \ll \sqrt{t}
\end{equation}
where  $c_0 = \int_{-\infty}^{+\infty} f(y) \mathrm{d} y = 2.913\dots$. Now, the mean cluster number is given by deriving $\left<  \ell(b) \right>$ with respect to $b$~\eqref{eq:Ncellc}, so only the subleading term $\propto C(b)  t^{-\frac12 - \eta}$ contributes, and we obtain the following:
\begin{equation}\label{eq:dense1-main}
   \left< N_c(b) \right> \sim 
   \begin{dcases} 
   b^{2\eta - 1} t^{-\frac12 - \eta} &  1 \ll b \ll \sqrt{t}  \\ 
    b^{-2\eta - 1} t^{-\frac12 - \eta} &  1 \gg b \gg 1/\sqrt{t}
   \end{dcases} \,.
\end{equation}
Since the above averages include extinct realizations, it is helpful to restrict to the non-extinct ones, recalling that the BBM survives with probability $\sim 1/t$~\eqref{eq:Pn}: %
\begin{equation}\label{eq:dense1-main}
   \boxed{  \left< \ell(b)  \right>_{\text{non-ex}} \simeq c_0 t^{\frac12}  \,,\,c_0 = 2.913\dots,  \frac1{\sqrt{t}} \ll b \ll \sqrt{t} \,,\; \left< N_c(b) \right>_{\text{non-ex}} \sim 
   \begin{dcases} 
   b^{2\eta - 1} t^{\frac12 - \eta} &  1 \ll b \ll \sqrt{t} \\ 
    b^{-2\eta - 1} t^{\frac12 - \eta} &  \frac1{\sqrt{t}} \ll b \ll 1
   \end{dcases} \,. }
\end{equation}
This is the main result of this section. The part of it concerning $N_c(b)$ is announced in \eqref{eq:Ncb_intro} in terms of the fractal dimension $D_{\mathrm{f}}= 1-2\eta$~\eqref{eq:etaDf}, where we also restored the dependence on $\beta$ and $D$ by requiring that $N_c(b)$ be dimensionless. 


A striking point of \eqref{eq:dense1-main} is that the leading asymptotic of $\left< \ell(b)  \right>_{\text{non-ex}} \sim t^{1/2} $ is independent of $b$, including the prefactor. This means that the particles form a rather compact bulk within the diffusive length scale $|x| \lesssim \sqrt{t}$, densely populating a finite portion of that length. Changing the coarse-graining scale affects that portion only by an infinitesimal amount in the $t\to\infty$ limit. Furthermore, the average number of clusters grows qualitatively more slowly than the extension, indicating the existence of large clusters. All of the above is reminiscent of the spatial distribution of all the positions visited by a critical branching {fractional} Brownian motion in certain dimensions~\cite{cao22}. 

We remark that $t F(x,t)$ is the probability, conditioned on non extinction, that there is at least one particle in $[x-b/2,x+b/2]$. Therefore, \eqref{eq:Ffinal} implies that, when $t$ is large, this probability is given by $f(x / \sqrt{t})$, plotted in Figure~\ref{fig:fandphi}. Similarly, $t \partial_b F(x,t)$ is the probability density (conditioned on non extinction) that the closest particle to $x$ is at distance $b/2$ from $x$. For $t$ large, this probability is proportional to $\partial_b C(b) t^{-\eta }\varphi(x/\sqrt{t})$, where $C(b)$ has the two limiting behaviors given in \eqref{eq:Ffinal}. The shape of $\varphi(y)$ (Figure~\ref{fig:fandphi}) with two maxima away from $y = 0$ means that it is more probable to have gaps away from the origin (which is where the BBM started). 


\subsection{Conditioned regimes ($\mu > 0$)}\label{sec:mu}
We now turn to study the averages with particle number conditioning, by setting $\mu > 0$ but $\mu \ll 1$. We also focus uniquely on the large gaps $b\gg1$. There are thus two large time scales:
\begin{enumerate}
\item $b^2$ is the characteristic time for a Brownian particle to displace by a distance $b$; thus, gaps of size $b$ can probably appear when and only when $t \gtrsim b^2$. In terms of the KPP equation, $b^2$ is the time it takes for the plateau to ``melt''  under diffusion.
\item $ \mu^{-1}$ is a time scale set by the conditioning on the particle number. Indeed, the effect of the chemical potential $\mu$ is to suppress the contribution of  the realizations with $n \gg \mu^{-1}$. Thus, by \eqref{eq:Pn}, for $t \ll \mu^{-1}$, there are virtually no such realizations, so $\mu$ has no effect; for $t \gg \mu^{-1}$, the typical realizations (with particle number $\gg \mu^{-1}$) are suppressed, and we are favoring realizations with atypically few particles.

In terms of the KPP equation, $\mu^{-1}$ is the time scale of the plateau melting under the effect of the nonlinear term. Indeed, If we solve the KPP equation \eqref{eq:KPPF} approximately by ignoring the diffusive term, we get an ODE $\dot{F} = -F^2$ which has a solution 
    \begin{equation} \label{eq:Fnonlinear}
       F(x,t) = \begin{dcases}
        \frac1{1 + t} & |x| < b \\ F_0 \approx \frac1{\mu^{-1} + t} & |x| > b 
        \end{dcases} \,.
    \end{equation}
Here we also approximated $(1-e^{-\mu}) \approx \mu$ as $\mu \ll 1$. The ratio between the two values tends to $1$ when $t \gg \mu^{-1}$. 
\end{enumerate}
From the above discussion, it is reasonable to expect that the asymptotic behavior of the solution to the KPP equation depends crucially on whether $\mu^{-1}$ or $b^2$ is greater; We will call the regime where $b^2 \ll \mu^{-1}$ ($b^2 \gg \mu^{-1}$) the typical gap regime (rare gap regime, respectively). The analysis of the typical gap regime (Section~\ref{sec:dense}) relies on the preceding $\mu = 0$ results; meanwhile, the rare gap regime (Section~\ref{sec:dilute}) calls for a distinct approach, somehow reminiscent of that in~\cite{ramola}. 
 
\subsubsection{The typical gap regime}\label{sec:dense}
In the typical gap regime where $b^2 \ll \mu^{-1} $, the plateau of the initial condition \eqref{eq:KPPF} will first melt by diffusion, when $t \sim b^2$. After that, there are two time regimes. At intermediate time, $b^2 \ll t \ll \mu^{-1}$, the solution can be viewed, for all practical purposes, as identical to the one with $\mu = 0$. That has been studied above, and we found the following asymptotic form:
\begin{equation}\label{eq:Ansatz_recall}
     F(x,t) = \frac1t \left[  f(x / \sqrt{t}) + C t^{-\eta} \varphi(x / \sqrt{t}) + \dots \right] \,,\, C \sim b^{2\eta}  \,,\, b^2 \ll t \ll \mu^{-1} \,.
\end{equation}
Now, when $t \gg \mu^{-1}$, the nonzero value of $\mu$ can no longer be ignored, and dictates the large distance asymptotics of the solution $F(x\to \infty) = F_0 \sim 1/t$. Thus, the Ansatz above with $f(|y| \to \infty) \to 0$ is invalid. 

Nevertheless, a similar Ansatz { with a different leading order scaling function $f(y) \to f_1(y) = 1$ becomes valid. Indeed, $f_1(y) = 1$ is a solution to \eqref{eq:ODEf}. Then, the eigenvalue equation \eqref{eq:H} becomes equivalent to that of an oscillator: the ground state energy is $\eta \to \eta_1 = 3/2$, and the ground state wavefunction $\varphi(y) \to \varphi_1(y) = e^{-y^2/4}$}. However, it is important to note that $ \varphi_1(y) $ is the dominant (with the slowest decay) perturbation \textit{that vanishes as }$|y| \to \infty$. There is one more perturbation, which is constant in $y$: $\varphi_0(y) = 1$, and with $\eta_0 = 1$. In terms of the Schr\"odinger Hamiltonian, it corresponds to an un-normalizable  wavefunction. Such a perturbation was not allowed {\blue for $\mu = 0$} above since $F$ has to vanish at infinity there. When $\mu > 0$, it must exist, and its amplitude is determined by the large-$x$ behavior of $F$: 
$$ F(x\to\infty) = F_0 = \frac1{\mu^{-1} + t} = \frac1t (1 - \mu^{-1} t^{-1} + \dots) \,, $$
see \eqref{eq:Flimit} above.

Gathering the above, we have 
\begin{equation} \label{eq:Ansatz2}
    F(x,t) = \frac{1}t \left[ 1 - \mu^{-1} t^{-1} + C_1 t^{-\frac32} e^{-x^2/{4t}} + \dots \right] \,,\,  b^2 \ll \mu^{-1} \ll  t
\end{equation}
Again, we need to {determine} the prefactor $C_1$. In principle, we need to characterize the crossover between the two asymptotic Ans\"atze; this is done in Appendix~\ref{sec:crossover}. However, the result of that analysis can be mostly recovered by a simple argument, which consists in matching the above two Ans\"atzes with $x= 0$ at the crossover time $t = \mu^{-1}$:
\begin{equation}
     C_1 \mu^{3/2} \sim f(0) + c_1 b^{2\eta} \mu^{\eta}  
\end{equation} 
This fixes $C_1$ and gives 
\begin{equation}\label{eq:dense2-F}
   F(x,t) = \frac{1}t \left[ 1  - \mu^{-1} t^{-1}  + \left( c_0' (t\mu)^{-\frac32}  +  c_1 t^{-\frac32} b^{2\eta} \mu^{\eta - \frac32} \right)  e^{-\frac{x^2}{4t}} + \dots \right]
\end{equation}
where $c_0' \approx f(0)$~\footnote{This is a rough approximation. A numerically exact estimate is $c_0' \approx 1$, see appendix \ref{sec:crossover}. 
} and $c_1$ is a order one constant independent of $b, t$ and $\mu$. Plugging that into \eqref{eq:ellbF} and \eqref{eq:NcbF} we have
\begin{equation} \label{eq:dense2-mu}
    \left< \ell(b) e^{-\mu n} \right> \sim t^{-2} \mu^{- \frac32}  \,,\, \left< N_c(b) e^{-\mu n} \right> \sim t^{-2} \mu^{\eta - \frac32} b^{2\eta - 1} \,.
\end{equation}
By comparing to \eqref{eq:On} and \eqref{eq:Omu}, we see that this corresponds to the following average conditioned on particle number: 
\begin{equation}\label{eq:dense2-n}
     \boxed{ \left< \ell(b)  \right>_n \sim  n^{\frac12}  \,,\, \left< N_c(b)  \right>_n \sim  n^{\frac12 - \eta} b^{2\eta - 1}  \,,\, 1\ll b \ll \sqrt{n}  \ll \sqrt{t} \,. }
\end{equation}
Eq.~\eqref{eq:dense2-n} is the $b \ll \sqrt{n}$ case of the result \eqref{eq:Ncbn_intro}. 

It is interesting to remark that \eqref{eq:dense2-n} is identical to \eqref{eq:dense1-main} upon replacing $n$ by $t$. This can be understood as follows: when the particle number $n \ll t$ is atypically small, the particles at $t$ have a recent common ancestor at time $t' = (t - n)$. As far as $\ell(b)$ and $N_c(b)$ are concerned, the effective ``age'' of the BBM is $n$ instead of ${t}$. Thus, the cluster structure discussed below \eqref{eq:dense1-main} prevail as we condition on a large particle number $n$. 

\subsubsection{The rare gap regime}\label{sec:dilute} 
In the rare gap regime, where $b^2 \gg \mu^{-1}$, the plateau of the initial condition in \eqref{eq:KPPF} will melt first from {the action of the nonlinear term in the KPP equation}. Indeed, \eqref{eq:Fnonlinear} is a good approximation at $t \ll b^2$ almost everywhere \footnote{except near the edges of the plateau. Indeed, the edge is softened by the diffusion term. It has a width $\sim \sqrt{t}$, much smaller than $b$.}. As a consequence, as $t$ exceeds $\mu^{-1}$, $F(x,t)$ becomes almost a constant. Indeed, the ratio between the values $F$ inside and far from the plateau is
\begin{equation}\label{eq:Fratio}
   \frac{ F(|x| \ll b)}{F_0} = \frac{t + \mu^{-1}}{t + 1} \approx 1 + \frac{1}{t \mu} \,.
\end{equation}
The last approximation is valid for $t \gg 1,  \mu^{-1} \gg 1$. Thus, we can fix a time $t_0 $ such that $t_0 \ll b^2$ (so the approximation \eqref{eq:Fnonlinear} is still valid) and that $ 1 / (\mu t_0)$ is small. At later time, the solution can only become closer and closer to the constant $F_0$, since both terms $\partial_x^2 F$ and $-F^2$ make the solution more uniform. Therefore, we can treat the KPP equation for $t \ge t_0$ by a perturbation expansion around the constant solution $F_0 = 1/(\mu^{-1} + t)$~\eqref{eq:F0}. The expansion is controlled by the small parameter $1 / (\mu t_0)$. We write 
\begin{equation}
    F = F_0 + a F_1 + a^2 F_2 + \dots  \,,\, \text{where } a = (\mu t_0)^{-1} \label{eq:Fseries} 
\end{equation}
is the small parameter. As we explained above, the zero-th order solution is the constant $F_0=1/(\mu^{-1} + t)$~\eqref{eq:F0}. Plugging \eqref{eq:Fseries} into \eqref{eq:KPPF} and comparing order by order in $a$, we find 
\begin{align}
    &(\partial_t - \partial_x^2 + 2 F_0) F_1 = 0  \,,\, F_1\vert_{t = t_0} =  F_0 \, \theta(b/2 - |x|) =  \frac{1}{\mu^{-1} + t_0} \theta(b/2 - |x|) \,, \label{eq:KPPF1}  \\
    &(\partial_t - \partial_x^2 + 2 F_0) F_2 = -F_1^2  \,,\, F_2\vert_{t = t_0} =  0 \,, \label{eq:KPPF2} 
\end{align}
and so on. Here, the initial condition of $F_1$ comes from \eqref{eq:Fratio}, which implies $F(|x|\ll b, t= t_0) \approx F_0 (1 + a)$ (while outside the plateau, $F(t = t_0) \approx F_0$). We will then extract the extension and gap distribution order by order:
\begin{align}
  &  \left< \ell (b) e^{-\mu n} \right> = \ell^{(1)} + \ell^{(2)} + \dots \,,\quad \ell^{(m)} := a^m \int  F_m(x,t) \mathrm{d} x \,, \label{eq:ellcbpert} \\
   &  \left< N_c (b) e^{-\mu n} \right> = N_c^{(1)} + N_c^{(2)} + \dots \,,\quad N_c^{(m)} := \partial_b  \ell^{(m)} \,,\label{eq:Ncbpert}
\end{align}

To carry out the perturbation calculations, it is convenient to introduce a retarded propagator $G$ defined by the following 
\begin{equation}
    (\partial_t - \partial_{xx} + 2F_0) G(x,t|y,s) =\delta(x-y)\delta(t-s)  \,,\, G\vert_{t < s} = 0\,.
\end{equation}
whose solution can be explicitly found
\begin{equation}
    G(x,t|y,s) = p(x-y, t-s) \frac{(\mu^{-1}+s)^2}{(\mu^{-1} +t)^2} \,,\, 
    p(z, u) = \frac{1}{\sqrt{4\pi u}} e^{-\frac{z^2}{4 u}} \,. \label{eq:green}
\end{equation}
Here $p$ is the usual diffusive kernel. Note that the Green function is translation invariant in space but not in time.

Using the Green function, we can readily solve the perturbative KPPs, \eqref{eq:KPPF1} and \eqref{eq:KPPF2}. At first order, we have 
\begin{equation}
    F_1(x,t) =  \int G(x, t| y, t_0) F_1(y, t_0)  \mathrm{d} y = \frac{(\mu^{-1} + t_0)}{(\mu^{-1} + t)^2} \int_{-b/2}^{b/2} p(x-y, t - t_0) \mathrm{d} y \label{eq:F1} \,.
\end{equation}
Plugging this into \eqref{eq:ellcbpert} and \eqref{eq:Ncbpert}, the integral over $p$ being unity, we find rather trivial geometrical information:
\begin{equation}
   \ell^{(1)} \sim  \frac{b}{\mu t^2}  \,,\,  N_c^{(1)} \sim \frac{1}{\mu t^2} \,. \label{eq:Ncborder1}
\end{equation} 
These can be understood as follows. The coarse-grain scale $b$ is so large that a realization will either have a single cluster of length $b$ or no clusters (if it is extinct). So a realization with $n > 0$ particles contributes $e^{-\mu n}$ (and $e^{-\mu n} b$) to  $ N_c^{(1)} $ (and $\ell^{(1)}$, respectively); extinct ones do not contribute. We can check that \eqref{eq:Ncborder1} follows from \eqref{eq:Pn} in the limit $t \gg \mu^{-1} \gg 1$: indeed, for the cluster number, \eqref{eq:Pn} implies $\left< e^{-\mu n} \mathbf{1}_{n>0} \right> \approx t^{-2}\int_0^\infty e^{-\mu n - n /t} \mathrm{d} n = 1/(t (1 + \mu t)) \approx 1 / (\mu t^2)$ (a similar calculation applies to the extension).

At second order, 
\begin{equation}
       F_2(x,t) =  - \int_{t_0}^{t} \mathrm{d} s \int  \mathrm{d} y    G(x,t|y,s) F_1(y, s)^2 \,. \label{eq:F2}
\end{equation}
The computation of $\ell^{(2)}$ is a bit lengthy, see Appendix~\ref{app:pert}. The result is simple when $t \gg b^2$:
\begin{equation}\label{eq:ell2}
    \ell^{(2)} \simeq - \frac{1}{\mu^2 t^2} \frac4b  \,,\, N_c^{(2)} \simeq   \frac{1}{\mu^2 t^2} \frac4{b^2} \,,\, 
\end{equation}
The higher order contributions $\ell^{(k)}$ and $ N_c^{(k)}$ are more tedious to compute, but their $t, \mu$ and $b$ dependence can be obtained by a diagrammatic power-counting (see Appendix~\ref{eq:diagram}):
\begin{equation}
    \ell^{(k)} \sim \frac{b^{3-2k}}{t^2 \mu^k} \,,\,  N_c^{(k)} \sim \frac{b^{2-2k}}{t^2 \mu^k } \,. \label{eq:higherorder}
\end{equation}
Since we are interested in the regime $b^2 \gg \mu^{-1}$, these contributions are more and more subleading as the order increases. Thus, it is justified to stop the perturbation expansion at second order (by contrast, in the $b^2 \ll \mu$, the present perturbation theory would break down). Combining both orders, we obtain
\begin{equation}\label{eq:dilute-mu}
     \left< N_c(b) e^{-\mu n} \right> = N_c^{(1)} + N_c^{(2)} + \dots \sim \frac{1}{\mu  t^2} + \frac{4}{\mu^2 t^2 b^2} + \dots  \,,\, 1 \ll\mu^{-1} \ll b^2 \ll t \,. 
\end{equation}
One might check, by \eqref{eq:On} and \eqref{eq:Omu}, that \eqref{eq:dilute-mu} is equivalent to the following gap distribution conditioned on the number of particles:
\begin{equation} \label{eq:dilutemain}
    \boxed{\left< N_c(b) \right>_n -1 \simeq \frac{4 n}{b^2}  \,,\, 1 \ll \sqrt{n} \ll b \ll \sqrt{t} \,.}
\end{equation}
Sending $t \to \infty$, we obtain the $b \gg \sqrt{n}$ case of the result \eqref{eq:Ncbn_intro} announced in the introduction. As we discussed there, eq.~\eqref{eq:dilutemain} is in full agreement with the universal gap distribution prediction~\eqref{eq:gap3_intro} of \cite{ramola,ramola15}, including the exact prefactor, as we discussed below \eqref{eq:Ncbn_intro}. In fact, the perturbative approach here is closely related to with the hierarchy of traveling wave equations considered in these works, which allows to obtain detailed information on the $k$-th gap. The order of expansion here corresponds to the number of particles. It is thus not surprising that we need to go to second order to access information about gaps.

\section{Numerical study}\label{sec:num}
We studied the model with two numerical methods: direct simulation, and numerical integration of the KPP equations. 

\subsection{Direct simulation}
We measured directly the cluster number $\left< N_c(b) \right>$ in a discrete-time analogue of the critical BBM. The time evolution takes place by stroboscopic steps. In each step, each particle replicates itself (into 2 particles at the same position) with probability $1/2$, and is removed with probability $1/2$. Then, each of the remaining particles displaces by an independent Gaussian variable of zero mean and unity variance. This model is expected to be in the same universality class as the continuous-time critical BBM, while being simpler to simulate. We also applied an importance sampling technique to generate more realizations with large number of particles, see Appendix \ref{app:num1} for details.

\begin{figure}
    \centering
    \includegraphics[scale=.6]{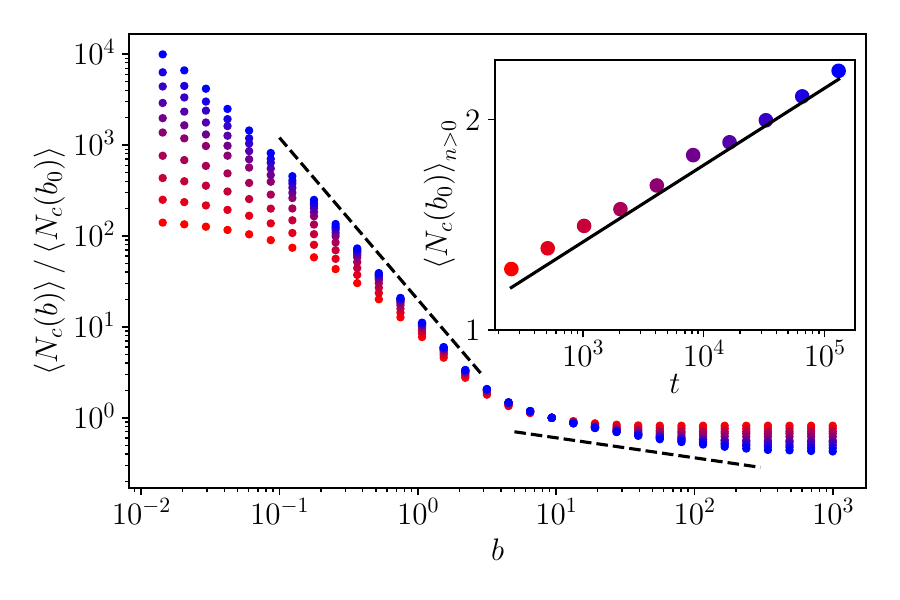}
    \caption{Results of the direct numerical simulation of a discrete-time analogue of the critical BBM.   \textit{Main}: the average cluster number $\left< N_c(b) \right>$ as a function of $b$, for various values of $t = 2^8, 2^9, \dots, 2^{17}$ (see inset for color code). The data are divided by $\left< N_c(b_0) \right> $, $b_0 = 5$ to highlight the $b$ dependence. The dashed lines indicate the predicted power laws $b^{-2\eta - 1} \approx b^{-1.78}$ (for small $b$) and $b^{2\eta - 1}\approx b^{-0.22}$ (for large $b$), with arbitrarily adjusted prefactors. 
    \textit{Inset}: The average cluster number with $b = b_0$ restricted to non-extinct realizations. The solid line indicates the predicted power law $t^{1/2-\eta} \approx t^{0.11}$,  with arbitrarily adjusted prefactors. See Appendix~\ref{app:num1} for numerical methods. We average over $10^6$ gaps for each value of $t$.}
    \label{fig:numerics}
\end{figure}

The main result of the simulation is presented in Figure~\ref{fig:numerics}. First, we plot the average cluster number, conditioned on non-extinction, with a coarse-graining scale of order unity ($b = b_0 = 5$), as a function of time. It shows a slow but visible increase, in nice agreement with the prediction $t^{1/2-\eta} \approx t^{0.11}$~\eqref{eq:dense1-main} with $\eta \approx 0.39$~\eqref{eq:eta-value}, see inset of Figure~\ref{fig:numerics}. Then, we focus on the $b$ dependence. As $t$ increases, we observe the emergence of two distinct power laws, with a crossover at $b \sim b_0$. The data are compatible with the predictions $b^{-2\eta - 1} \approx b^{-1.78}$ for small $b$ and $b^{2\eta - 1}\approx b^{-0.22}$ for large $b$. We note however that the relatively small value of the large-gap exponent makes it challenging to confirm without ambiguity; indeed, in Figure~\ref{fig:numerics}, the large-gap power law appears essentially flat compared to the small-gap one. We shall find more convincing evidence by numerically integrating the KPP equation.

\subsection{Integrating the KPP equation}
We have seen that precisely observing the large gap exponent in \eqref{eq:dense1-main}  requires many orders of magnitude in space and time. In fact, it remains laborious even if we resort to the semi-numerical approach of integrating the KPP equation. To overcome this difficulty we used a integrate-and-coarse-grain scheme that allows us to access large space-time scales with moderate computation resource, see Appendix \ref{app:num2} for details. 

The main results are displayed in Figure~\ref{fig:numerics1_kpp}. In the left panel, we plotted the average cluster number, for large $b$, in the same way as in Figure~\ref{fig:numerics} above, i.e., factoring away the dependence on $t$. By going up to $t \sim 10^8$, we observe a well-established power law $\left< N_c(b) \right> \sim b^{2\eta - 1}$, further corroborating the result \eqref{eq:dense1-main}. 

We also integrated the KPP equation with $\mu \ne 0$, in order to observe the crossover between the typical and rare gap regimes. More precisely, we test the predictions \eqref{eq:dense2-mu} and \eqref{eq:dilute-mu}, which can be reformulated as follows: in the long time limit, 
\begin{equation} \label{eq:Ncbmu-num}
    -b\partial_b \left< N_c(b) e^{-\mu n} \right> t^2 \mu \sim \begin{dcases}
    (b \mu^{\frac12})^{2\eta - 1} & 1 \ll b \ll \mu^{-\frac12} \\
     8 (b  \mu^{\frac12} )^{-2} & b \gg  \mu^{-\frac12}
    \end{dcases} \,.
\end{equation}
where the prefactor in the $b \gg  \mu^{-\frac12}$ is exact, and the $b\partial_b$ derivative is introduced to get rid of the $b$-independent term in \eqref{eq:dilute-mu}. This prediction is nicely confirmed by numerics, see Figure~\ref{fig:numerics1_kpp}, right panel. This result complements the numerical work in \cite{ramola,ramola15} performed on fixed particle number sectors. 

\begin{figure}
    \centering
   \includegraphics[width=0.48\textwidth]{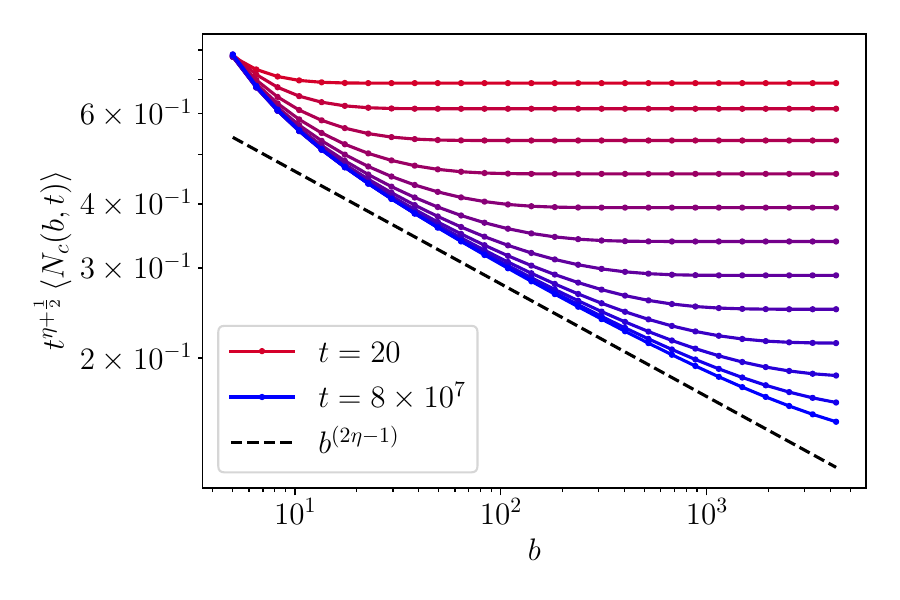} \includegraphics[width=0.48\textwidth]{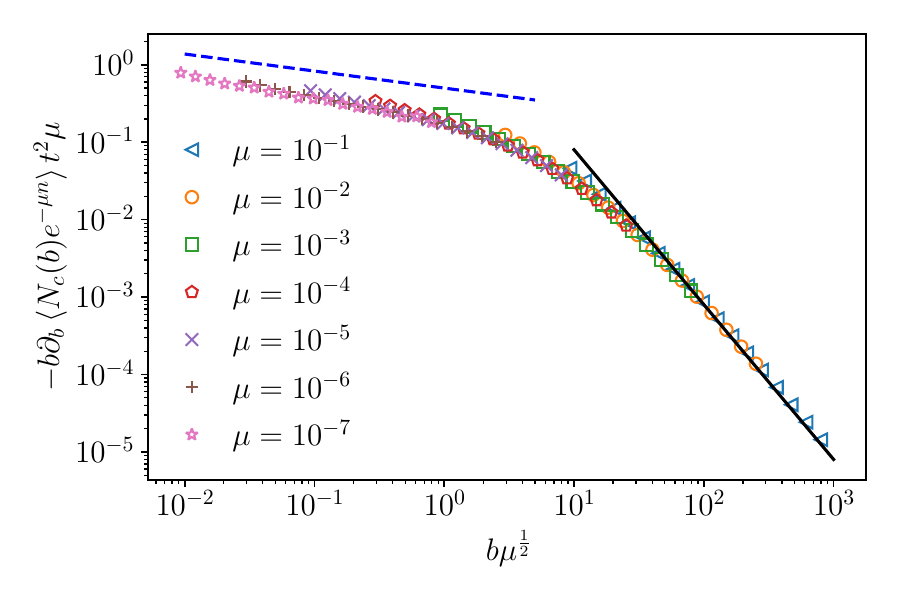}
    \caption{Results obtained from numerical integration of the KPP equation. \textit{Left panel}: 
    Average cluster number, with the predicted $t$ dependence removed, $t^{\eta + 1/2}\left< N_c(b) \right>$, for $b \in [5, 5000]$ and $t = 20, \dots, 8 \times 10^7$ (represented by color, from red to blue). The dashed line indicates the $b^{2\eta - 1}$ ($\eta = 0.39$) prediction, with arbitrarily adjusted prefactor. 
        \textit{Right panel}: Testing the prediction~\eqref{eq:Ncbmu-num} on the cluster number with particle number conditioning, for a few values of $\mu$. The value of $t$ alternates between $3.3\times 10^8$ and $1.3\times 10^9$. The blue dashed line plots the power laws $(b \mu^{1/2})^{2\eta-1}$ with an arbitrarily adjusted prefactor. The black solid lines depicts $ 8 (b \mu^{1/2})^{-2}$ with the predicted exact prefactor. See Appendix~\ref{app:num2} for numerical methods.}
    \label{fig:numerics1_kpp}
\end{figure}

\section{Discussion}\label{sec:conclude}
We characterized the clustering structure of the 1D critical branching Brownian motion, adopting an approach introduced in \cite{cao22}. This allowed us reconcile the result of \cite{ramola,ramola15} focusing on gap statistics in fixed particle number sectors, and that of \cite{mytnik17} focusing on the continuum limit. Our approach is based on a mapping to the KPP equation. Its asymptotic solution is always given by a perturbative expansion, yet around different backgrounds depending on the regime. Our analysis of the KPP equation is not rigorous, but is well supported by numerics. We note that the analysis of the KPP equation ---in particular in the regime $t \ll \mu^{-1}$, corresponding to no conditioning on the particle number --- is reminiscent of that of the instanton equation in our previous work~\cite{cao22}. There, we analyzed the clusters of all the positions (integrated over time) visited by a critical branching fractional Brownian motion. The number of these clusters is also governed by a nontrivial exponent controling the subleading asymptotics of the solution, analogous to the exponent $\eta$ (and to the fractal dimension of the boundary $D_{\mathrm{f}} = 1-2\eta$) crucial to the present work.

Critical branching Brownian motion displays clustering (patchiness) in higher dimensions as well. The results of this work apply to the projection of a $d$-dimension BBM onto one direction. We can also consider generalizing the present method to $d > 1$, by assigning a patch (a disk in 2D, a ball in 3D, \dots) of radius $b$ to each point, so that nearby balls overlap and form clusters. A similar mapping to a KPP equation allows us to compute directly the avarage $d$-dimensional volume $V$ occupied by the patches. However, differentiating with respect to $b$ no longer yields the cluster number $N_c(b)$ for $d>1$ but other topological numbers. For example, in 2D, the second derivative of the area $\partial_b^2 \mathcal{A}$ is proportional to the difference between the number of clusters and holes. Therefore, counting clusters in higher dimensions remains an interesting problem.  On the other hand, we expect it to be rather straightforward to extend the present approach to 1D branching  \textit{fractional} Brownian motion, i.e., random walks with long-range hopping; results on this topic will be reported elsewhere.

\begin{acknowledgements}
We thank Satya Majumdar for useful discussions, in particular for pointing out the Reference~\cite{ramola}. {P. Le Doussal acknowledges support from ANR grant ANR-17-CE30-0027-01 RaMaTraF.}
\end{acknowledgements}

\appendix 
\section{Derivation of the KPP equation}\label{app:kpp}
We derive of the KPP equation~\eqref{eq:kpp} by a backward recursion argument.  The initial condition follows from definition. Now, the backward recursion consists in considering what can happen during the initial elapse $t \in [0, \mathrm{d} t]$:
\begin{enumerate}
    \item The initial particle branches into two with probability $ \mathrm{d} t$. In that case $E(x) \to E(x)^2$ (because from that moment, the two individuals act independently from now on with the same law), thus giving a contribution $ (E^2 - E) \mathrm{d} t$ to $\mathrm{d} E$.
    \item The initial particle is annihilated with probability $ \mathrm{d} t$. Then $E(x) \to 1$, giving a contribution $ ( 1 - E)\mathrm{d} t $ to  $\mathrm{d} E$. 
    \item The initial particle displaces to $\mathrm{d} x_0$, then $E(x) \to E(x - \mathrm{d} x_0)$, giving a contribution 
    $$ E(x - \mathrm{d} x_0) - E(x) = - \partial_x E  \left< \mathrm{d} x_0 \right> +  \frac12 \partial_x^2 E \left< (\mathrm{d} x_0)^2 \right> =  \partial_x^2 E \mathrm{d t} $$
    to $\mathrm{d} E$: here we used Ito calculus (expanding to second order) and that $ \left< \mathrm{d} x_0 \right> = 0, \left< (\mathrm{d} x_0)^2 \right> = 1$. 
\end{enumerate}
Combining all the contributions, we obtain \eqref{eq:kpp}.

\section{About the dense (super-Brownian motion) limit}\label{sec:SBM}

The super-Brownian motion (SBM) can be defined as a dense limit of the critical BBM. A common way to proceed (see e.g.~\cite{perkins1,slade2002scaling,pierrenew})
 is to consider a large parameter $M \gg 1$, and perform the following steps:
\begin{enumerate}
    \item Set the branching/annihilation rate to be large $\beta = M$ (while keeping the diffusion constant $D$ of order one).
    \item Modify the initial condition, and start with $n(0) \propto M$ particles at the origin. 
     \item Define the local density as 
    \begin{equation}
        \rho(x,t) := \frac1M \sum_{i=1}^{n(t)} \delta(x - x_i(t)) \,,
    \end{equation} where $n(t)$ is the number of particles at time $t$ and $x_i(t), i = 1, \dots, n(t)$ their positions. 
\end{enumerate}
Then, it is known that the cumulant generating function of $\rho(x,t)$ satisfies a KPP equation with a delta-function initial condition. This is the basic framework of \cite{mytnik17}. The goal of this Appendix is to {discuss that only step 1 above is essential for the observables studied here.}

To see why the initial condition is not essential, consider the critical BBM with $\beta = M$ (step 1) but only a single particle initially (no step 2). Then, consider what happens after a small lapse $M^{-1} \ll  t \ll 1$. Since $M t \gg 1$, if the process has not gone extinct, it will have $\sim Mt$ particles. Since $t \ll 1$ and $D$ is of order one, these particles have no time to diffuse and thus are still infinitesimally close to the origin. Therefore, we have shown that, conditioned on non-extinction after an infinitesimal time, starting with one particle is equivalent to starting with $O(M)$ particles at the same position. 
{We note that the procedure of restricting to non-extinct realizations is often achieved in another way, by rescaling the probability measure by $M$. That gives rise to a un-normalized measure, called the ``canonical measure''~\cite{perkins1,slade2002scaling}. In the context of mean-field theory of avalanches (Brownian force model), the canonical measure corresponds to studying the avalanche response with respect to an infinitesimal kick, see \cite{pierrenew}, around eq.~(25) therein.}

Step 3 is a choice of observable. The ones considered in this work are different; fundamentally, they are all related to the probability that $[x - b/2, x + b/2]$ contains at least one particle at time $t$, denoted $F(x,t)$ [see \eqref{eq:Ext} above with $\mu = 0$]. If we consider a single initial particle at $x=0$, $F(x,t)$ satisfies a KPP equation, which is the following with dimensions restored:
\begin{equation}\label{eq:KPPSBM}
    \partial_t F = D \partial_x^2 F - \beta F^2 \,,\, F(x, t=0) = \theta(b/2-|x|) \,.
\end{equation}
Now, in the SBM limit $\beta = M \gg 1$, we can consider $F = \tilde{F} / M$ so that 
\begin{equation}
    \partial_t \tilde{F} = D \partial_x^2  \tilde{F} -  \tilde{F}^2 \,,\, \tilde{F}(x, t=0) = M \theta(b/2-|x|) \,.
\end{equation}
Initially, the nonlinear term dominates with respect to the diffusion one, so that 
\begin{equation}
     \tilde{F} \approx \frac{1}{M^{-1} + t} \theta(b/2-|x|) 
\end{equation}
This approximation is valid until $\tilde{F}$ becomes of order unity. Observe that the RHS of the above equation has a limit as $M\to \infty$ (uniformly in $t \in [t_0, \infty)$ for any fixed $t_0 > 0$). Therefore, we may conclude that in the SBM limit, the solution to \eqref{eq:KPPSBM} has the following form:
\begin{equation} \label{eq:Frescale}
    F(x,t)= \frac1M \tilde{F}(x,t)
\end{equation}
where $\tilde{F}(x,t)$ satisfies a KPP equation with a diverging initial condition
\begin{equation} \label{eq:FtildeKPP}
    \partial_t \tilde{F} = D \partial_x^2  \tilde{F} -  \tilde{F}^2 \,,\, \tilde{F}(t\to0) \simeq \frac1t \theta(b/2-|x|) \,.
\end{equation}
To interpret these formulas, recall that $1/M$ is the probability that the process is not extinct at $t_0 = 1$. Hence, by \eqref{eq:Frescale}, for $t > t_0$, $\tilde{F}(x,t)$ is the probability that $[x-b/2, x+b/2]$ is not empty at $t$, conditioned on non-extinction at an early time. (As per the discussion above, $\tilde{F}(x,t)$ is also the probability of $[x-b/2, x+b/2]$ not being empty at $t$ with an initial condition of $O(M)$ particles near the origin.) Therefore, the observables we study in this work have a well-defined behavior in the SBM limit. This limit can be directly studied via the KPP equation \eqref{eq:FtildeKPP}. In this work, we took the alternative approach: we focus on the BBM, and then discuss the SBM limit of the results, see \eqref{eq:NcbSBM} in Section \ref{sec:res_typ}.

\section{Crossover between the two asymptotic Ans\"atze}\label{sec:crossover}
\begin{figure}
    \centering
    \includegraphics[scale=0.5]{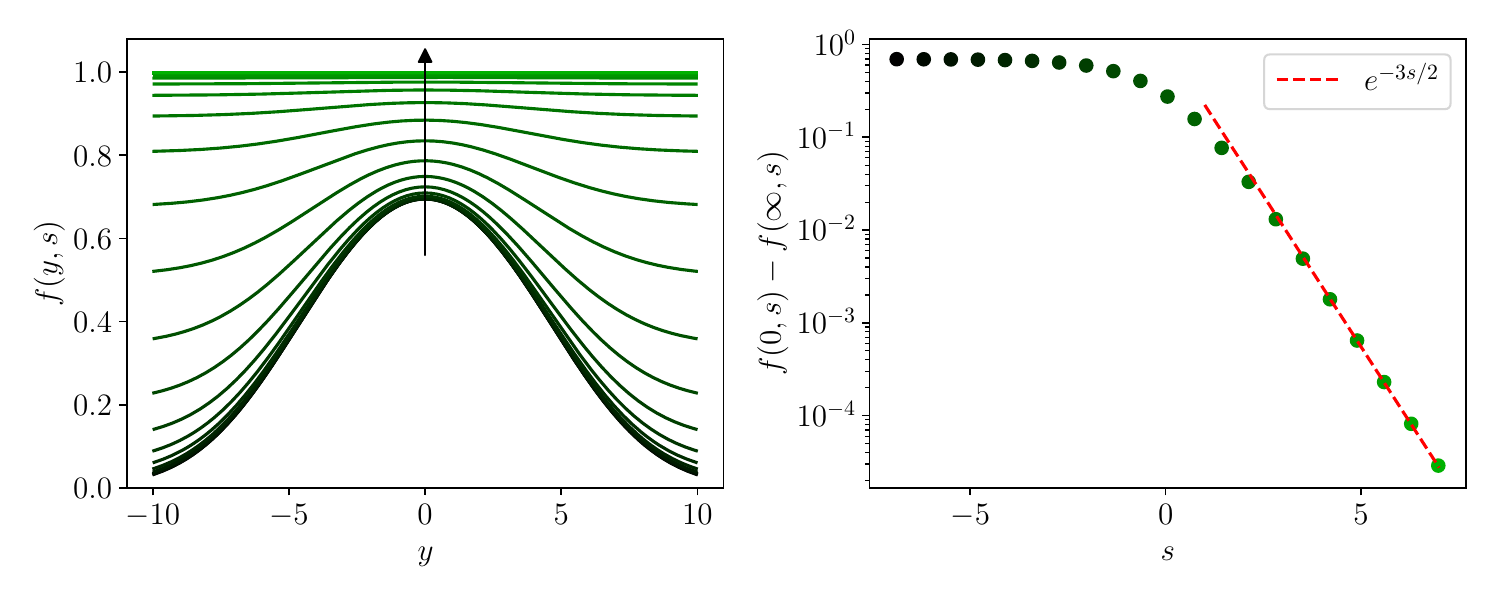}
    \caption{Crossover between the two asymptotic Ans\"atze. The left panel plots the solution to \eqref{eq:fgeneral} that interpolates between the $t \ll \mu^{-1}$ Ansatz \eqref{eq:Ansatz_recall} and the $t\gg \mu^{-1}$ one \eqref{eq:Ansatz2}, as a function of $y = x/\sqrt{t}$, for $s = \ln (t \mu^{-1}) \in [-7,7]$ (from bottom to top, also indicated by color, see right panel for color code). The right panel plots the $s$ dependence of the center, $f(y =0,s)$, from which we subtracted $f(y=\infty, s) = t / (\mu^{-1} + t) = e^{s} / (1 + e^s).$ It decays as $e^{-3s/2}$, with a prefactor $A_1 \approx 1.02$ (best fit, red dashed line). This provides an estimate of the amplitude $A_1$ in \eqref{eq:fsplus}, which is equal to $c_0'$ in \eqref{eq:dense2-F}.}
    \label{fig:crossover}
\end{figure}
In this appendix we characterize in more detail the crossover between the $t \ll \mu^{-1}$ and $t \gg \mu^{-1}$ asymptotic Ans\"atze discussed in Section~\ref{sec:dense}. 

\subsection{Leading term}
We first focus on the crossover between leading-order terms. For this, we perform a change of variables, 
\begin{equation}\label{eq:Ff}
    F(x,t) = \frac1t f(y,s) \,,\, \text{where } y = x/\sqrt{t}, s = \ln (t\mu) \,.
\end{equation}
Here and below, to avoid confusion, we will write $f(y,s)$ to refer to the time-dependent function just defined, while $f(y)$ stands for the leading term of the Ansatz~\eqref{eq:Ansatz}. Then the KPP equation $\partial_t F = \partial_{x}^2 F - F^2$ is equivalent to 
\begin{equation}\label{eq:fgeneral}
    \partial_s f(y,s) = \partial_y^2 f(y,s)+ \frac{y}2 \partial_y f(y,s)+ f(y,s) - f(y,s)^2  \,.
\end{equation}
This is the time-dependent generalization of \eqref{eq:ODEf} above. The leading terms of the two Ans\"atze --- $f(y)$ in \eqref{eq:Ansatz_recall} (plotted in Figure~\ref{fig:fandphi}), and $f_1(y) = 1$ in \eqref{eq:Ansatz2} --- are time-independent solutions of \eqref{eq:fgeneral}. The crossover {between them} is described by a time-dependent solution, which tends to $f(y)$ as $s \to -\infty$ and to $f_1(y) = 1$ as $s\to \infty$ (note that the rescaled time $s = \ln(\mu t)$ is the logarithm of the physical time, so $s \to - \infty$ and $s \to \infty$ correspond to $t \ll \mu^{-1}$ and $t \gg \mu^{-1}$, respectively). Also, its large-distance behavior is also fixed as $ f(|y| \to\infty) = {t} F_0  = {t}/(\mu^{-1} + t) = {e^s}/(1 + e^s)$, see \eqref{eq:Flimit} above. Numerical integration of the KPP equation (the original one, not the rescaled one studied here; see Appendix~\ref{app:kpp} for methods) indicates that the crossover solution with the above boundary conditions is unique. Indeed, we observe that solutions to the KPP equation with different values of $\mu$ coincide with each other in the re-scaled coordinate system $(y, s, f)$, as long as $1 \lesssim b^2 \ll \mu^{-1}$. In Figure~\ref{fig:crossover}, we plot the crossover solution obtained numerically. 

Let us characterize the limiting behaviors of the crossover solution:
\begin{itemize}
    \item At large distances 
\begin{equation}\label{eq:flargey}
    f(|y| \to \infty, s) \simeq \frac{e^s}{1 + e^s} \,,
\end{equation}
which increases from $0$ to $1$ as $s$ increases from $-\infty$ to $+\infty$. 
\item As $s \to -\infty$, $f(y,s)$ tends to $f(y)$ of \eqref{eq:Ansatz_recall} and \eqref{eq:Ansatz}, plotted in Figure~\ref{fig:fandphi}. Linearizing \eqref{eq:fgeneral} around that fixed point, we find that the difference $\delta f(y,s) = f(y,s) - f(y)$ satisfies the time-dependent analogue of \eqref{eq:H} above, 
\begin{equation}\label{eq:Hrecall}
    \partial_s (\delta f) = -H(\delta f) \,,\,  H = -\partial_y^2 - \frac{y}2 \partial_y + 2 f(y) - 1 \,.
\end{equation}
We checked that $H$ has an eigenfunction $\psi(y)$ with a negative eigenvalue  $-1$, corresponding to an instability. This should be distinguished from the eigenvalue $\eta > 0$ which corresponds to a decaying eigenfunction, see Section~\ref{sec:mu0delta} above.
A plot of $\psi(y)$ can be found in Figure~\ref{fig:psi}. In particular, $\psi(|y|\to \infty)$ tends to a nonzero constant; hence, it corresponds to a non-normalizable wavefunction of the Schr\"odinger Hamiltonian~\eqref{eq:HS}. The $s \to -\infty$ asymptotics of $\delta f$ is dominated by this unstable mode:  
\begin{equation}
    f(y,s \to -\infty) = f(y) + e^{s} \psi(y) + \dots \,,\, \psi(|y|\to\infty) \simeq 1 \,.
\end{equation}
Note that the $|y|\to\infty$ limit is in agreement with \eqref{eq:flargey}.

\item As {$s \to + \infty$}, $f(y,s)$ tends to $f_1(y) = 1$. Linearization around this fixed point leads to \eqref{eq:Hrecall} with the Hamiltonian replaced by 
\begin{equation}
    H = -\partial_y^2 - \frac{y}2 \partial_y + 1 \,. \label{eq:Hoscillator_recall}
\end{equation} Its leading eigenfunctions are the constant  $\varphi_0(y) = 1$ with eigenvalue $1$, and the Gaussian $\varphi_1(y) = e^{-y^2/4}$ with eigenvalue $3/2$. Therefore, we have  
\begin{equation}\label{eq:fsplus}
    f(y,s \to +\infty) = 1 + A_0 e^{-s} \varphi_0(y) + A_1 e^{-3s/2} \varphi_1(y)  + \dots =  1 - e^{-s}  + A_1 e^{-3s/2} e^{-y^2 / 4}  + \dots \,.
\end{equation}
Here, the amplitude $A_0 = -1$ has been fixed by comparing with \eqref{eq:flargey}. We estimated numerically that $A_1 \approx 1.02$, see \ref{fig:crossover} (right panel). Note that $A_1$ is equal to the coefficient $c_0'$ in \eqref{eq:dense2-F}. 
\end{itemize}

\begin{figure}
    \centering
    \includegraphics[scale=0.5]{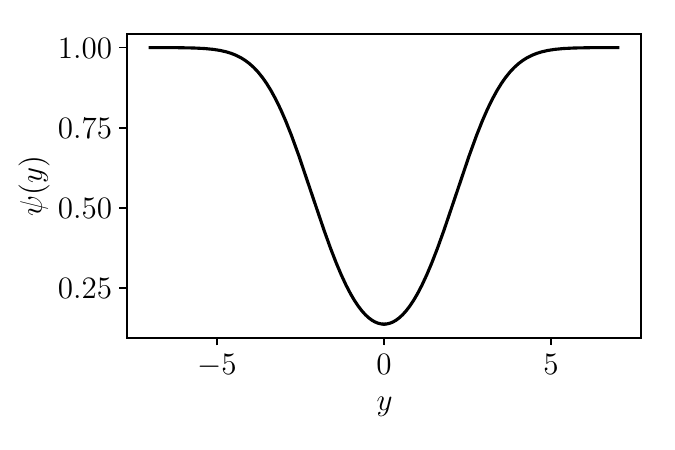}
    \caption{The unstable eigenfunction $\psi(y)$ of the Hamiltonian $H$ \eqref{eq:Hrecall} with eigenvalue $-1$. It tends to $1$ as $|y|\to \infty$.  }
    \label{fig:psi}
\end{figure}

\subsection{Subleading term}
\begin{figure}
    \centering
    \includegraphics[scale=0.5]{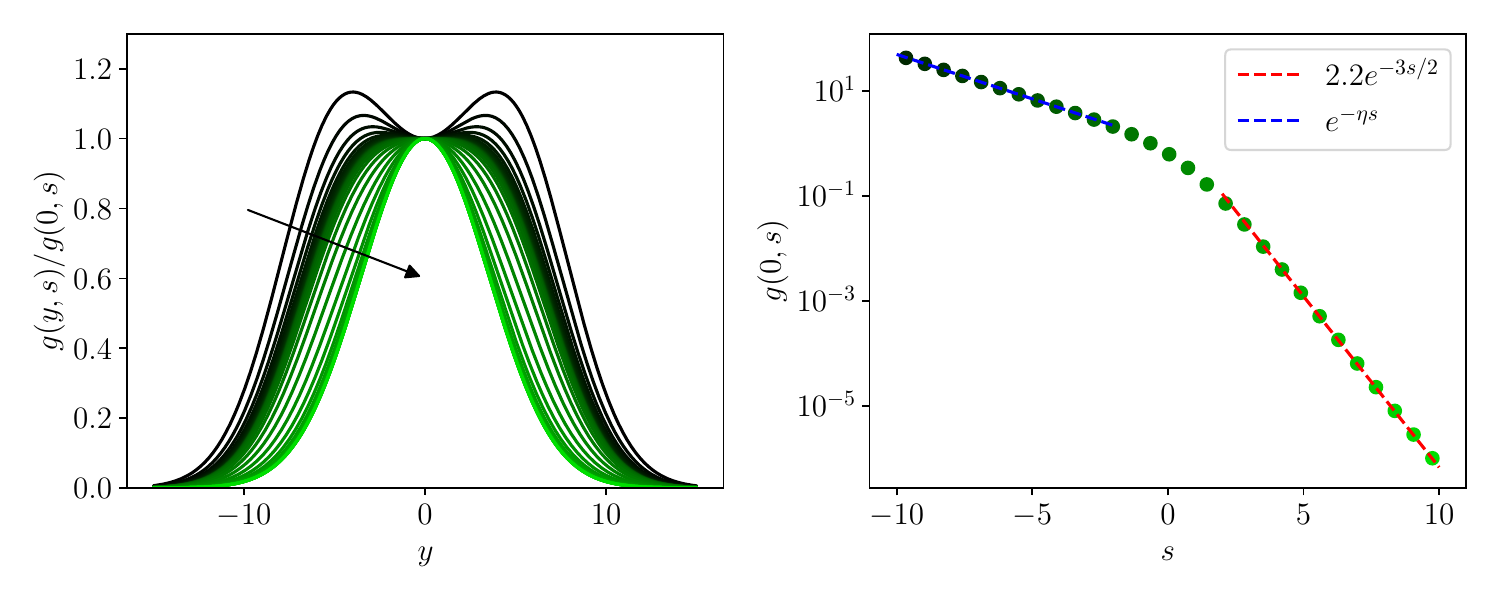}
    \caption{Crossover between the subleading term of the two asymptotic Ans\"atze. The left panel plots the solution to \eqref{eq:timedependent} with the asymptotic condition \eqref{eq:gasym}, as a function of $y$, for $s = \ln (t \mu^{-1}) \in [-10, 10]$ (indicated by color, see right panel for color code). The right panel plots the $s$ dependence of the center, $g(y = 0,s)$. It decays as $e^{-3s/2}$ for $s \to \infty$, with a prefactor $A_g \approx 2.2$ (best fit, red dashed line).}
    \label{fig:crossover1}
\end{figure}
We now consider a small perturbation around the leading-order crossover solution studied above, that is, we replace $f(y,s) \to f(y,s) + \varepsilon g(y,s)$ in \eqref{eq:Ff} and \eqref{eq:fgeneral}, where $\varepsilon$ is a small parameter (to be fixed below). 
This leads to a linearized equation for $g$:
\begin{equation}\label{eq:timedependent}
     \partial_s g(y,s) = - H(s) g(y,s) \,,\, H(s) := -\partial_y^2 - \frac{y}2 \partial_y + 2 f(y,s) - 1 \,.
\end{equation}
It has a time-dependent ``Hamiltonian'' that interpolates between \eqref{eq:Hrecall} and \eqref{eq:Hoscillator_recall}. A unique solution is specified by the asymptotic condition 
\begin{equation}\label{eq:gasym}
    g(y,s) \simeq e^{-\eta s} \varphi(y) \,, s \to -\infty \,.
\end{equation}
where $\eta \approx 0.39$ and $\varphi$ are the leading decaying eigenvalue and eigenfunction of \eqref{eq:Hrecall} [see below \eqref{eq:H} for further discussion, and Figure~\ref{fig:fandphi} for a plot.] The full solution can be again extracted from the numerical integration of the KPP equation, and is plotted in Figure~\ref{fig:crossover1}. As $s \to \infty$, it has the following asmyptotics:
\begin{equation} \label{eq:gsplus}
    g(y,s) \simeq A_g e^{-3s/2} e^{-y^2/4} \,, s \to +\infty \,,\,
\end{equation}
where we estimated $A_g \approx 2.2$. 

We now fix the small parameter $\varepsilon$. For this, we match the $t \ll \mu^{-1}$ (or $s\to-\infty$) behavior of the  crossover solution with perturbation
\begin{equation}\label{eq:Ffg}
     F(x,t) = \frac1t \left[f(x/\sqrt{t}, \ln (\mu t)) + \varepsilon g(x, \ln (\mu t)) \right] = \frac1t  \left[f(x/\sqrt{t}) + \varepsilon (\mu t)^{-\eta} \varphi(x/\sqrt{t}) + \dots \right]
\end{equation}   
with the Ansatz \eqref{eq:Ansatz_recall}, which is 
$$F(x,t) = \frac1t  \left[f(x/\sqrt{t}) + C t^{-\eta} \varphi(x/\sqrt{t}) \right] \,,\, C \simeq A_c b^{2\eta} \,.  $$
($A_c$ is a constant independent of $b, t, \mu$). As a result, we find $\varepsilon = C \mu^{\eta} \sim b^{2\eta} \mu^{\eta}$. 
This, combined with \eqref{eq:fsplus} and \eqref{eq:gsplus}, implies the following $t \gg \mu^{-1}$ (or $s \to +\infty$) asymptotics:
\begin{align}
    F(x,t) =& \frac1t \left[f(x/\sqrt{t}, \ln (\mu t)) + \varepsilon g(x, \ln (\mu t)) \right]  \nonumber \\
    = &\frac1t \left[1 - (\mu t)^{-1} + \left( A_1 (\mu t)^{-3/2}  + A_g A_c  t^{-3/2} b^{2\eta} \mu^{\eta-\frac32}  \right) e^{-\frac{x^2}{4t}} + \dots  \right]
\end{align}
This is exactly \eqref{eq:dense2-F}; in particular, the coefficients in \eqref{eq:dense2-F} $c_0' = A_1$ and $c_1 = A_g A_c$ are indeed order one constants that do not depend on $b, t$ and $\mu$.

In conclusion, a more elaborate analysis of the crossover between the two Ans\"atze confirms the simple matching argument presented in Section~\ref{sec:dense}. 

\section{Perturbation calculations}
\subsection{Second order}\label{app:pert}
In this appendix, we derive \eqref{eq:ell2}. We will work in the regime $1 \ll \mu^{-1} \ll t_0 \ll b^2 \ll t$ and will always use them to simplify expressions. First, by \eqref{eq:F2} and \eqref{eq:ellcbpert}, 
\begin{align}
    \ell^{(2)} & \sim - a^2 \int_{t_0}^{t} \mathrm{d} s  \int \mathrm{d} x \mathrm{d} y  G(x,t|y,s)  F_1(y, s)^2  \\ \sim
    &  - a^2 \int_{t_0}^{t} \mathrm{d} s \frac{s^2}{t^2} \int \mathrm{d}  x \mathrm{d} y \, p(x-y, t-s) F_1(y, s)^2  \\
    =  & - a^2 \int_{t_0}^{t} \mathrm{d} s \frac{s^2}{t^2} \int \mathrm{d} y F_1(y, s)^2
\end{align}
where in the last line, we used the property $\int p(z,u) \mathrm{d} z = 1$ of the standard diffusion propagator. Now, plugging in \eqref{eq:F1}, 
\begin{align}
     \ell^{(2)} & \sim - \frac{1}{ \mu^2 t^2} \int_{t_0}^{t} \frac{ \mathrm{d} s}{s^2} \int \mathrm{d} y \int_{-\frac{b}2}^{\frac{b}2} \mathrm{d} u \int_{-\frac{b}2}^{\frac{b}2} \mathrm{d} v  p(y - u, s - t_0) p(y - v, s - t_0)  \\
    & =  - \frac{1}{\mu^2 t^2} \int_{t_0}^{t} \frac{ \mathrm{d} s}{s^2} \left[ 
    b \, \mathrm{erf}\left(\frac{b}{2 \sqrt{2 (s-t_0) }}\right)- \sqrt{\frac{8 (s-t_0) }{\pi }}   \left(1 - e^{-\frac{b^2}{8 (s-t_0) }}\right) \right] \\
    & \sim - \frac{1}{ \mu^2 t^2 t_0} \left[ \sqrt{2 \pi t_0} e^{\frac{b^2}{8 t_0}} \left(1-\text{erf}\left(\frac{b}{2 \sqrt{2} \sqrt{t_0}}\right)\right)+b-\sqrt{2 \pi t_0} \right]
\end{align}
The integral of the second line is exact (recall $\mathrm{erf}(x) = \frac2{\sqrt{\pi}} \int^x_0 e^{-z^2} \mathrm{d} z$), and the last line is exact for $t \gg b^2$. We can further simplify using $t_0 \ll b^2$:
\begin{equation}
       \ell^{(2)} \sim  - \frac{1}{ \mu^2 t^2} \frac4b -\frac{1}{ \mu^2 t^2} \frac{b}{t_0} \,.
\end{equation}
The second term here is similar to $\ell^{(1)}$ \eqref{eq:Ncborder1}, but is $a = 1 / (\mu t_0)$ times smaller. Dismissing it gives us the desired result \eqref{eq:ell2}.

\subsection{Higher orders: diagrammatic expansion and power-counting}\label{eq:diagram}
The perturbation expansion of $\left< \ell(b) \right>$ in Section~\ref{sec:dilute} can be represented by diagrams, see Fig.~\ref{fig:diagrams}. This diagrammatic expansion is closely related to that of the Brownian force model of mean-field avalanches~\cite{pldw12,pldw13}. The diagrams at order $k$ are trees with one root, $k-1$ internal vertices, and $k$ leaves. The Feynmann rules for calculating the contribution of a diagram to $\ell^{(k)}$ are the following:
\begin{figure}
    \centering
    \includegraphics[scale=.6]{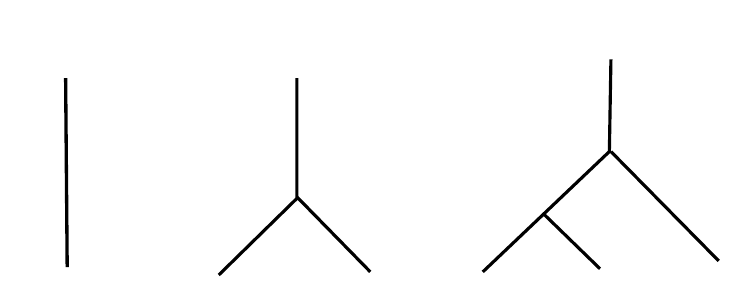}
    \caption{Diagrammatic representations of the perturbation expansion at order $1$ (left), $2$ (center) and $3$ (right, plus its mirror image). }
    \label{fig:diagrams}
\end{figure}
\begin{enumerate}
    \item To each vertex (i.e., internal vertices, root and leaves) is associated a space-time point.
    \item To each edge is associated with one retarded propagator ~\eqref{eq:green} 
    $$ G(x,t|y,s) = \frac{(\mu^{-1} + s)^2}{(\mu^{-1} + t)^2} p(x-y,t-s)  \theta(t-s) \,,\, p(z, u) = \frac{e^{\frac{z^2}{4\pi u}}}{\sqrt{4\pi u}} $$
    between the vertices it connects. The later time $t$ corresponds to the vertex closer to the root (upper vertex in Figure~\ref{fig:diagrams}). We will assign the $(\mu^{-1} + s)^2$ to the vertex at time $s$ and $1/(\mu^{-1} + t)^2/\sqrt{t-s}$ to that at time $t$. 
    \item The root is integrated only over space $\in \R$, with time fixed to $t$.
    \item The leaves are integrated only over space $x \in [-b/2, b/2]$, with time fixed to $t_0 = 1/(a\mu) \gg \mu$. Moreover, to each leaf is associated with a factor $1/(\mu^{-1} + t_0)$, see \eqref{eq:KPPF1}.
    \item The internal vertices are integrated over both space $\in \R$ and time $\in \R$ (then constrained by the causality of the propagator).
    \item Finally, we multiply the result  by $a^{k}$, as per \eqref{eq:ellcbpert}.
\end{enumerate}
Now let us count the powers of $t, \mu$ and $b$ of a diagram. First, the root's integral can be performed independently of the rest of the diagram, giving a contribution $1/(\mu^{-1} + t^2) \sim 1/t^2$ (since $t \gg 1/\mu^{-1}$) from the numerator of the propagator. $t$ will not occur in the rest of the diagram, except as an upper time limit that will be sent to infinity (since we are interested in the $t \to\infty$ asymptotics). So the power of $t$ is simply $1/t^2$ for any diagram. 

Next, $\mu$ can only occur in the numerator $(\mu^{-1} + t_0)^2$ of the Green function connecting a leaf (rule 2), and the factor $1/(\mu^{-1} + t_0)$ associated to it (rule 4). Since there are $k$ leaves, we have a factor $(\mu^{-1} + t_0)^k \sim t_0^k \sim a^{-k} \mu^{-k}$ since $t_0 \gg \mu^{-1}$. Combined with the $a^k$ factor (rule 6), we have a $\mu^{-k}$ power: this is the $\mu$ dependence. By the way, we see that the $a$-dependence has been cancelled: the result of the calculation has a well-defined limit as $a \to 0$. 

Finally, we count the powers of $b$. For this, we count the length $[x]$ and time $[t]$ powers that remain, i.e., those associate with the space-integral of the leaves, and those associated with the internal vertices (i.e., excluding leaves and the root). Each leaf-integral gives a $[x]$ power, which adds to $[x^k]$ (there being $k$ leaves), by rule 4. For each internal vertex, it has a $[x t]$ power because of its space-time integral, by rule 5. The three propagator above it (see Figure~\ref{fig:diagrams}) contributes $[t^2]$, and the two below it contribute $([t^{-1/2}] [t^{-2}])^2$, by rule 2. Converting $[t] = [x^2]$ (by diffusive scaling of the propagator), we obtain $[xt][t^2]([t^{-1/2}] [t^{-2}])^2 = [x t^{-2}] = [x^{-3}]$ per internal vertex. Since there are $k-1$ vertices, we have $[x^{-3k+3}]$ for all internal vertices. Adding both leaf and internal vertex contributions, we get $[x^{-2k + 3}]$, which translates to a $b^{-2k+3}$ dependence on $b$.

In summary, we have shown that every diagram of $\ell^{(k)}$ contributes a power \begin{equation}
    t^{-2} \mu^{-k} b^{-2k+3} \,,
\end{equation}
as announced in \eqref{eq:higherorder} above.

\section{Numerical methods}
\subsection{Importance sampling}\label{app:num1}
The idea is to sample the realization with a statistical weight biased by the factor $e^{ \mu n(T)}$, where $n(T)$ is the number of particle at the maximal time step $T$ of the simulation, and $\mu > 0$ is the {\blue bias parameter}, chosen positive to favor realizations with large number of particles. {It should be neither too small (no effect) nor too large (causes divergence); in practice $\mu = 1/t$ is a good compromise}. This bias is then corrected by an inverse weight {\blue $ e^{- \mu n(T)}$} when computing observables. 

Thanks to the tree structure of the BBM model, the weighed ensemble can be sampled directly. To see how, consider the partition function 
\begin{equation}
     Z(\mu, T) := \left< e^{ \mu n(T)} \right> \,.
\end{equation} 
By definition, the importance sampling consists in generating configurations of the statistical ensemble defined by $Z$. By a backward recursion argument, it is not hard to show 
\begin{equation}
     Z(\mu, T) = \frac12 (1 + Z(\mu, T-1)^2) \,,\, T > 0 \,,\, 
\end{equation}
and $ Z(\mu, 0) = e^{\mu}$. Using this recursion relation we can easily compute numerically $Z(\mu, t)$ for $t = 0, 1, \dots, T$. Then, a moment of thought shows that, to sample $Z(\mu, T)$, we should simulate the branching process with a sequence of modified weights: at the $k$-th step, each particle branches with probability $1/(2 Z(\mu, T - k +1))$ and annihilates with probability  $Z(\mu, T-k)/(2 Z(\mu, T - k +1))$ (we use the sequence of partition functions \textit{backwards} in time, since we work with a backward recursion). The random-walk part of the simulation is not affected by the importance sampling.

\subsection{Integration and coarse-grain scheme}\label{app:num2}
The numerical integration of the KPP equation involves an integrate-and-coarse-grain scheme, reminiscent of a real-space renormalization group, in order to reach large time and space with moderate resource. Indeed, we expect the solution to becomes smoother and smoother in time. Therefore, we can gradually coarse-grain the solution, discarding short-distance information, without losing precision. This is done by iteratively applying the rescaling $t = \alpha^2 \tilde{t}$,  $x = \alpha \tilde{x}, F = \alpha^{-2} \tilde{F}$ \eqref{eq:rescaling}, under which the KPP equation is invariant.

More concretely, we approximate the space by a lattice with finite size $[-A/2, A/2]$ and spacing $\epsilon$. $F$ is represented by an array $N = A / \epsilon$ data points. We execute the following steps:
\begin{enumerate}
    \item \textit{Initial step}. Integrate the KPP equation from the initial condition for a duration of $t_0$. 
    \item \textit{Coarse-grain}. We remove every other data points of the solution just obtained (so $\alpha = 2$), multiply it by $4$, and pad both ends of the array by the edge value repeated $N/4$ times: 
    \begin{equation}
        (F_1, \dots, F_N) \leftarrow (\underbrace{F_1, \dots, F_1}_{\text{$N/4$ times}}, F_1, F_3, F_5 \dots, F_{N-1}, \underbrace{ F_{N-1}, \dots, F_{N-1}}_{\text{$N/4$ times}})
    \end{equation}
    As a result, the lattice remains the same but it now represents the rescaled $\tilde{x}$ variable.
    Note that the time is now rescaled to $t_0 / 4$.
    \item \textit{Integrate}. Evolve the solution by the KPP equation for a duration of $t_0 - t_0/4$ so that the (rescaled) time is again $t_0$.
\end{enumerate}
We can then carry on by looping steps 2 and 3 indefinitely. After $k$ iterations, the algorithm outputs the rescaled {\blue solution} $\tilde{F}(\tilde{x}, \tilde{t})$ at $\tilde{t} = t_0$. It is related to the physical solution in original coordinates $F = \tilde{F} 4^{-k}$, $x = \tilde{x} 2^k$, $t = t_0 4^k$. Hence, we obtain the solution at a space-time scale that is an exponential of the computation time; the memory cost is constant. The geometric quantities are calculated using \eqref{eq:ellbF} and \eqref{eq:NcbF}. {To handle the $\partial_b$ derivative involved in the cluster number \eqref{eq:NcbF}, we integrate an auxiliary system for $G= \partial_b F$ along with $F$ using the same techniques. Integrating $G$ gives the cluster number directly.}

For the scheme to be accurate, the time $t_0$ should be large enough: the solution at $t= t_0$ should be sufficient smooth so that the coarse-grain step does not erase too much information.  The lattice is then chosen to afford a good approximation of the solution for $t \in [0, t_0]$. In practice, a moderate set of parameters $t_0 = 20$, $\epsilon = 0.25$ and $A = 10^3$ is adequate for our purposes. The integration of the KPP equation is performed by a first-order Trotterization scheme (with $\delta t = 0.1$), in which the diffusion and nonlinear terms are treated in momentum and real space, respectively.

\bibliography{ref}

\begin{thebibliography}{25}%
\makeatletter
\providecommand \@ifxundefined [1]{%
 \@ifx{#1\undefined}
}%
\providecommand \@ifnum [1]{%
 \ifnum #1\expandafter \@firstoftwo
 \else \expandafter \@secondoftwo
 \fi
}%
\providecommand \@ifx [1]{%
 \ifx #1\expandafter \@firstoftwo
 \else \expandafter \@secondoftwo
 \fi
}%
\providecommand \natexlab [1]{#1}%
\providecommand \enquote  [1]{``#1''}%
\providecommand \bibnamefont  [1]{#1}%
\providecommand \bibfnamefont [1]{#1}%
\providecommand \citenamefont [1]{#1}%
\providecommand \href@noop [0]{\@secondoftwo}%
\providecommand \href [0]{\begingroup \@sanitize@url \@href}%
\providecommand \@href[1]{\@@startlink{#1}\@@href}%
\providecommand \@@href[1]{\endgroup#1\@@endlink}%
\providecommand \@sanitize@url [0]{\catcode `\\12\catcode `\$12\catcode
  `\&12\catcode `\#12\catcode `\^12\catcode `\_12\catcode `\%12\relax}%
\providecommand \@@startlink[1]{}%
\providecommand \@@endlink[0]{}%
\providecommand \url  [0]{\begingroup\@sanitize@url \@url }%
\providecommand \@url [1]{\endgroup\@href {#1}{\urlprefix }}%
\providecommand \urlprefix  [0]{URL }%
\providecommand \Eprint [0]{\href }%
\providecommand \doibase [0]{https://doi.org/}%
\providecommand \selectlanguage [0]{\@gobble}%
\providecommand \bibinfo  [0]{\@secondoftwo}%
\providecommand \bibfield  [0]{\@secondoftwo}%
\providecommand \translation [1]{[#1]}%
\providecommand \BibitemOpen [0]{}%
\providecommand \bibitemStop [0]{}%
\providecommand \bibitemNoStop [0]{.\EOS\space}%
\providecommand \EOS [0]{\spacefactor3000\relax}%
\providecommand \BibitemShut  [1]{\csname bibitem#1\endcsname}%
\let\auto@bib@innerbib\@empty
\bibitem [{\citenamefont {Zhang}\ \emph {et~al.}(1990)\citenamefont {Zhang},
  \citenamefont {Serva},\ and\ \citenamefont {Polikarpov}}]{zhang90}%
  \BibitemOpen
  \bibfield  {author} {\bibinfo {author} {\bibfnamefont {Y.~C.}\ \bibnamefont
  {Zhang}}, \bibinfo {author} {\bibfnamefont {M.}~\bibnamefont {Serva}},\ and\
  \bibinfo {author} {\bibfnamefont {M.}~\bibnamefont {Polikarpov}},\ }\bibfield
   {title} {\bibinfo {title} {Diffusion reproduction processes},\ }\href
  {https://doi.org/10.1007/BF01026554} {\bibfield  {journal} {\bibinfo
  {journal} {Journal of Statistical Physics}\ }\textbf {\bibinfo {volume}
  {58}},\ \bibinfo {pages} {849} (\bibinfo {year} {1990})}\BibitemShut
  {NoStop}%
\bibitem [{\citenamefont {Tsimring}\ \emph {et~al.}(1996)\citenamefont
  {Tsimring}, \citenamefont {Levine},\ and\ \citenamefont
  {Kessler}}]{tsimring}%
  \BibitemOpen
  \bibfield  {author} {\bibinfo {author} {\bibfnamefont {L.~S.}\ \bibnamefont
  {Tsimring}}, \bibinfo {author} {\bibfnamefont {H.}~\bibnamefont {Levine}},\
  and\ \bibinfo {author} {\bibfnamefont {D.~A.}\ \bibnamefont {Kessler}},\
  }\bibfield  {title} {\bibinfo {title} {{RNA Virus Evolution via a
  Fitness-Space Model}},\ }\href {https://doi.org/10.1103/PhysRevLett.76.4440}
  {\bibfield  {journal} {\bibinfo  {journal} {Phys. Rev. Lett.}\ }\textbf
  {\bibinfo {volume} {76}},\ \bibinfo {pages} {4440} (\bibinfo {year}
  {1996})}\BibitemShut {NoStop}%
\bibitem [{\citenamefont {Houchmandzadeh}(2002)}]{Houchmandzadeh0}%
  \BibitemOpen
  \bibfield  {author} {\bibinfo {author} {\bibfnamefont {B.}~\bibnamefont
  {Houchmandzadeh}},\ }\bibfield  {title} {\bibinfo {title} {Clustering of
  diffusing organisms},\ }\href {https://doi.org/10.1103/PhysRevE.66.052902}
  {\bibfield  {journal} {\bibinfo  {journal} {Phys. Rev. E}\ }\textbf {\bibinfo
  {volume} {66}},\ \bibinfo {pages} {052902} (\bibinfo {year}
  {2002})}\BibitemShut {NoStop}%
\bibitem [{\citenamefont {Houchmandzadeh}(2008)}]{Houchmandzadeh}%
  \BibitemOpen
  \bibfield  {author} {\bibinfo {author} {\bibfnamefont {B.}~\bibnamefont
  {Houchmandzadeh}},\ }\bibfield  {title} {\bibinfo {title} {Neutral clustering
  in a simple experimental ecological community},\ }\href
  {https://doi.org/10.1103/PhysRevLett.101.078103} {\bibfield  {journal}
  {\bibinfo  {journal} {Phys. Rev. Lett.}\ }\textbf {\bibinfo {volume} {101}},\
  \bibinfo {pages} {078103} (\bibinfo {year} {2008})}\BibitemShut {NoStop}%
\bibitem [{\citenamefont {Houchmandzadeh}(2009)}]{Houchmandzadeh2}%
  \BibitemOpen
  \bibfield  {author} {\bibinfo {author} {\bibfnamefont {B.}~\bibnamefont
  {Houchmandzadeh}},\ }\bibfield  {title} {\bibinfo {title} {Theory of neutral
  clustering for growing populations},\ }\href
  {https://doi.org/10.1103/PhysRevE.80.051920} {\bibfield  {journal} {\bibinfo
  {journal} {Phys. Rev. E}\ }\textbf {\bibinfo {volume} {80}},\ \bibinfo
  {pages} {051920} (\bibinfo {year} {2009})}\BibitemShut {NoStop}%
\bibitem [{\citenamefont {Bailey}\ \emph {et~al.}(1975)\citenamefont {Bailey}
  \emph {et~al.}}]{bailey1975mathematical}%
  \BibitemOpen
  \bibfield  {author} {\bibinfo {author} {\bibfnamefont {N.~T.}\ \bibnamefont
  {Bailey}} \emph {et~al.},\ }\href@noop {} {\emph {\bibinfo {title} {The
  mathematical theory of infectious diseases and its applications}}}\ (\bibinfo
   {publisher} {Charles Griffin \& Company Ltd, 5a Crendon Street, High
  Wycombe, Bucks HP13 6LE.},\ \bibinfo {year} {1975})\BibitemShut {NoStop}%
\bibitem [{\citenamefont {Kendall}(1956)}]{kendall1956deterministic}%
  \BibitemOpen
  \bibfield  {author} {\bibinfo {author} {\bibfnamefont {D.~G.}\ \bibnamefont
  {Kendall}},\ }\bibfield  {title} {\bibinfo {title} {Deterministic and
  stochastic epidemics in closed populations},\ }in\ \href@noop {} {\emph
  {\bibinfo {booktitle} {Proceedings of the Third Berkeley Symposium on
  Mathematical Statistics and Probability, Volume 4: Contributions to Biology
  and Problems of Health}}}\ (\bibinfo {organization} {University of California
  Press},\ \bibinfo {year} {1956})\ pp.\ \bibinfo {pages}
  {149--165}\BibitemShut {NoStop}%
\bibitem [{\citenamefont {Dumonteil}\ \emph {et~al.}(2013)\citenamefont
  {Dumonteil}, \citenamefont {Majumdar}, \citenamefont {Rosso},\ and\
  \citenamefont {Zoia}}]{dumonteil13extent}%
  \BibitemOpen
  \bibfield  {author} {\bibinfo {author} {\bibfnamefont {E.}~\bibnamefont
  {Dumonteil}}, \bibinfo {author} {\bibfnamefont {S.~N.}\ \bibnamefont
  {Majumdar}}, \bibinfo {author} {\bibfnamefont {A.}~\bibnamefont {Rosso}},\
  and\ \bibinfo {author} {\bibfnamefont {A.}~\bibnamefont {Zoia}},\ }\bibfield
  {title} {\bibinfo {title} {Spatial extent of an outbreak in animal
  epidemics},\ }\href {https://doi.org/10.1073/pnas.1213237110} {\bibfield
  {journal} {\bibinfo  {journal} {Proceedings of the National Academy of
  Sciences}\ }\textbf {\bibinfo {volume} {110}},\ \bibinfo {pages} {4239}
  (\bibinfo {year} {2013})}\BibitemShut {NoStop}%
\bibitem [{\citenamefont {Meyer}\ \emph {et~al.}(1996)\citenamefont {Meyer},
  \citenamefont {Havlin},\ and\ \citenamefont {Bunde}}]{meyer96}%
  \BibitemOpen
  \bibfield  {author} {\bibinfo {author} {\bibfnamefont {M.}~\bibnamefont
  {Meyer}}, \bibinfo {author} {\bibfnamefont {S.}~\bibnamefont {Havlin}},\ and\
  \bibinfo {author} {\bibfnamefont {A.}~\bibnamefont {Bunde}},\ }\bibfield
  {title} {\bibinfo {title} {Clustering of independently diffusing individuals
  by birth and death processes},\ }\href
  {https://doi.org/10.1103/PhysRevE.54.5567} {\bibfield  {journal} {\bibinfo
  {journal} {Phys. Rev. E}\ }\textbf {\bibinfo {volume} {54}},\ \bibinfo
  {pages} {5567} (\bibinfo {year} {1996})}\BibitemShut {NoStop}%
\bibitem [{\citenamefont {Lawson}\ and\ \citenamefont
  {Jensen}(2007)}]{lawson07}%
  \BibitemOpen
  \bibfield  {author} {\bibinfo {author} {\bibfnamefont {D.~J.}\ \bibnamefont
  {Lawson}}\ and\ \bibinfo {author} {\bibfnamefont {H.~J.}\ \bibnamefont
  {Jensen}},\ }\bibfield  {title} {\bibinfo {title} {Neutral evolution in a
  biological population as diffusion in phenotype space: Reproduction with
  local mutation but without selection},\ }\href
  {https://doi.org/10.1103/PhysRevLett.98.098102} {\bibfield  {journal}
  {\bibinfo  {journal} {Phys. Rev. Lett.}\ }\textbf {\bibinfo {volume} {98}},\
  \bibinfo {pages} {098102} (\bibinfo {year} {2007})}\BibitemShut {NoStop}%
\bibitem [{\citenamefont {P{\'a}zsit}\ and\ \citenamefont
  {P{\'a}l}(2007)}]{pazsit2007neutron}%
  \BibitemOpen
  \bibfield  {author} {\bibinfo {author} {\bibfnamefont {I.}~\bibnamefont
  {P{\'a}zsit}}\ and\ \bibinfo {author} {\bibfnamefont {L.}~\bibnamefont
  {P{\'a}l}},\ }\href@noop {} {\emph {\bibinfo {title} {Neutron fluctuations: A
  treatise on the physics of branching processes}}}\ (\bibinfo  {publisher}
  {Elsevier},\ \bibinfo {year} {2007})\BibitemShut {NoStop}%
\bibitem [{\citenamefont {Zoia}\ \emph {et~al.}(2014)\citenamefont {Zoia},
  \citenamefont {Dumonteil}, \citenamefont {Mazzolo}, \citenamefont
  {de~Mulatier},\ and\ \citenamefont {Rosso}}]{deMulatier2014}%
  \BibitemOpen
  \bibfield  {author} {\bibinfo {author} {\bibfnamefont {A.}~\bibnamefont
  {Zoia}}, \bibinfo {author} {\bibfnamefont {E.}~\bibnamefont {Dumonteil}},
  \bibinfo {author} {\bibfnamefont {A.}~\bibnamefont {Mazzolo}}, \bibinfo
  {author} {\bibfnamefont {C.}~\bibnamefont {de~Mulatier}},\ and\ \bibinfo
  {author} {\bibfnamefont {A.}~\bibnamefont {Rosso}},\ }\bibfield  {title}
  {\bibinfo {title} {Clustering of branching brownian motions in confined
  geometries},\ }\href {https://doi.org/10.1103/PhysRevE.90.042118} {\bibfield
  {journal} {\bibinfo  {journal} {Phys. Rev. E}\ }\textbf {\bibinfo {volume}
  {90}},\ \bibinfo {pages} {042118} (\bibinfo {year} {2014})}\BibitemShut
  {NoStop}%
\bibitem [{\citenamefont {de~Mulatier}\ \emph {et~al.}(2015)\citenamefont
  {de~Mulatier}, \citenamefont {Dumonteil}, \citenamefont {Rosso},\ and\
  \citenamefont {Zoia}}]{deMulatier2015}%
  \BibitemOpen
  \bibfield  {author} {\bibinfo {author} {\bibfnamefont {C.}~\bibnamefont
  {de~Mulatier}}, \bibinfo {author} {\bibfnamefont {E.}~\bibnamefont
  {Dumonteil}}, \bibinfo {author} {\bibfnamefont {A.}~\bibnamefont {Rosso}},\
  and\ \bibinfo {author} {\bibfnamefont {A.}~\bibnamefont {Zoia}},\ }\bibfield
  {title} {\bibinfo {title} {The critical catastrophe revisited},\ }\href
  {https://doi.org/10.1088/1742-5468/2015/08/p08021} {\bibfield  {journal}
  {\bibinfo  {journal} {Journal of Statistical Mechanics: Theory and
  Experiment}\ }\textbf {\bibinfo {volume} {2015}},\ \bibinfo {pages} {P08021}
  (\bibinfo {year} {2015})}\BibitemShut {NoStop}%
\bibitem [{\citenamefont {Ramola}\ \emph {et~al.}(2014)\citenamefont {Ramola},
  \citenamefont {Majumdar},\ and\ \citenamefont {Schehr}}]{ramola}%
  \BibitemOpen
  \bibfield  {author} {\bibinfo {author} {\bibfnamefont {K.}~\bibnamefont
  {Ramola}}, \bibinfo {author} {\bibfnamefont {S.~N.}\ \bibnamefont
  {Majumdar}},\ and\ \bibinfo {author} {\bibfnamefont {G.}~\bibnamefont
  {Schehr}},\ }\bibfield  {title} {\bibinfo {title} {Universal order and gap
  statistics of critical branching brownian motion},\ }\href
  {https://doi.org/10.1103/PhysRevLett.112.210602} {\bibfield  {journal}
  {\bibinfo  {journal} {Phys. Rev. Lett.}\ }\textbf {\bibinfo {volume} {112}},\
  \bibinfo {pages} {210602} (\bibinfo {year} {2014})}\BibitemShut {NoStop}%
\bibitem [{\citenamefont {Ramola}\ \emph {et~al.}(2015)\citenamefont {Ramola},
  \citenamefont {Majumdar},\ and\ \citenamefont {Schehr}}]{ramola15}%
  \BibitemOpen
  \bibfield  {author} {\bibinfo {author} {\bibfnamefont {K.}~\bibnamefont
  {Ramola}}, \bibinfo {author} {\bibfnamefont {S.~N.}\ \bibnamefont
  {Majumdar}},\ and\ \bibinfo {author} {\bibfnamefont {G.}~\bibnamefont
  {Schehr}},\ }\bibfield  {title} {\bibinfo {title} {Branching brownian motion
  conditioned on particle numbers},\ }\href
  {https://doi.org/https://doi.org/10.1016/j.chaos.2014.12.013} {\bibfield
  {journal} {\bibinfo  {journal} {Chaos, Solitons \& Fractals}\ }\textbf
  {\bibinfo {volume} {74}},\ \bibinfo {pages} {79} (\bibinfo {year}
  {2015})}\BibitemShut {NoStop}%
\bibitem [{\citenamefont {Kolmogorov}\ \emph {et~al.}(1937)\citenamefont
  {Kolmogorov}, \citenamefont {Petrovsky},\ and\ \citenamefont
  {Piscounov}}]{kpp}%
  \BibitemOpen
  \bibfield  {author} {\bibinfo {author} {\bibfnamefont {A.}~\bibnamefont
  {Kolmogorov}}, \bibinfo {author} {\bibfnamefont {I.}~\bibnamefont
  {Petrovsky}},\ and\ \bibinfo {author} {\bibfnamefont {N.}~\bibnamefont
  {Piscounov}},\ }\bibfield  {title} {\bibinfo {title} {Etude de l'\'equation
  de la diffusion avec croissance de la quantit \'e de mati\`ere et son
  application \`a un probl\`eme biologique},\ }\href@noop {} {\bibfield
  {journal} {\bibinfo  {journal} {Bull. Univ. Etat Moscou A}\ }\textbf
  {\bibinfo {volume} {1}},\ \bibinfo {pages} {1} (\bibinfo {year}
  {1937})}\BibitemShut {NoStop}%
\bibitem [{per()}]{perkins1}%
  \BibitemOpen
  \href@noop {} {}\bibinfo {note} {Ed. Perkins, {\it Super-Brownian motion and
  critical spatial stochastic systems}, Can. Math. Bull., 47(2), 280 (2004),
  see also
  \url{http://www.math.ubc.ca/~perkins/superbrownianmotionandcriticalspatialsystems.pdf}}\BibitemShut
  {NoStop}%
\bibitem [{\citenamefont {Slade}(2002)}]{slade2002scaling}%
  \BibitemOpen
  \bibfield  {author} {\bibinfo {author} {\bibfnamefont {G.}~\bibnamefont
  {Slade}},\ }\bibfield  {title} {\bibinfo {title} {Scaling limits and
  super-brownian motion},\ }\href
  {https://www.ams.org/journals/notices/200209/fea-sladecolor.pdf?trk=200209fea-sladecolor&cat=collection}
  {\bibfield  {journal} {\bibinfo  {journal} {Notices AMS}\ }\textbf {\bibinfo
  {volume} {49}},\ \bibinfo {pages} {1056} (\bibinfo {year}
  {2002})}\BibitemShut {NoStop}%
\bibitem [{\citenamefont {Le~Doussal}(2022)}]{pierrenew}%
  \BibitemOpen
  \bibfield  {author} {\bibinfo {author} {\bibfnamefont {P.}~\bibnamefont
  {Le~Doussal}},\ }\bibfield  {title} {\bibinfo {title} {Equivalence of
  mean-field avalanches and branching diffusions: From the brownian force model
  to the super-brownian motion},\ }\href
  {http://iopscience.iop.org/article/10.1088/1751-8121/ac8d3b} {\bibfield
  {journal} {\bibinfo  {journal} {Journal of Physics A: Mathematical and
  Theoretical}\ } (\bibinfo {year} {2022})}\BibitemShut {NoStop}%
\bibitem [{\citenamefont {{Le Doussal}}\ and\ \citenamefont
  {Wiese}(2012)}]{pldw12}%
  \BibitemOpen
  \bibfield  {author} {\bibinfo {author} {\bibfnamefont {P.}~\bibnamefont {{Le
  Doussal}}}\ and\ \bibinfo {author} {\bibfnamefont {K.~J.}\ \bibnamefont
  {Wiese}},\ }\bibfield  {title} {\bibinfo {title} {Distribution of velocities
  in an avalanche},\ }\href {https://doi.org/10.1209/0295-5075/97/46004}
  {\bibfield  {journal} {\bibinfo  {journal} {{EPL} (Europhysics Letters)}\
  }\textbf {\bibinfo {volume} {97}},\ \bibinfo {pages} {46004} (\bibinfo {year}
  {2012})}\BibitemShut {NoStop}%
\bibitem [{\citenamefont {Le~Doussal}\ and\ \citenamefont
  {Wiese}(2013)}]{pldw13}%
  \BibitemOpen
  \bibfield  {author} {\bibinfo {author} {\bibfnamefont {P.}~\bibnamefont
  {Le~Doussal}}\ and\ \bibinfo {author} {\bibfnamefont {K.~J.}\ \bibnamefont
  {Wiese}},\ }\bibfield  {title} {\bibinfo {title} {Avalanche dynamics of
  elastic interfaces},\ }\href {https://doi.org/10.1103/PhysRevE.88.022106}
  {\bibfield  {journal} {\bibinfo  {journal} {Phys. Rev. E}\ }\textbf {\bibinfo
  {volume} {88}},\ \bibinfo {pages} {022106} (\bibinfo {year}
  {2013})}\BibitemShut {NoStop}%
\bibitem [{pie(2022)}]{pierrenewBFM}%
  \BibitemOpen
  \href@noop {} {\bibfield  {journal} {\bibinfo  {journal} {P. Le Doussal, {\it
  More on the Brownian force model: avalanche shapes, tip driven, higher d},
  ArXiv:2203.10544}\ } (\bibinfo {year} {2022})}\BibitemShut {NoStop}%
\bibitem [{\citenamefont {Mueller}\ \emph {et~al.}(2017)\citenamefont
  {Mueller}, \citenamefont {Mytnik},\ and\ \citenamefont {Perkins}}]{mytnik17}%
  \BibitemOpen
  \bibfield  {author} {\bibinfo {author} {\bibfnamefont {C.}~\bibnamefont
  {Mueller}}, \bibinfo {author} {\bibfnamefont {L.}~\bibnamefont {Mytnik}},\
  and\ \bibinfo {author} {\bibfnamefont {E.}~\bibnamefont {Perkins}},\
  }\bibfield  {title} {\bibinfo {title} {{On the boundary of the support of
  super-Brownian motion}},\ }\href {https://doi.org/10.1214/16-AOP1141}
  {\bibfield  {journal} {\bibinfo  {journal} {The Annals of Probability}\
  }\textbf {\bibinfo {volume} {45}},\ \bibinfo {pages} {3481 } (\bibinfo {year}
  {2017})}\BibitemShut {NoStop}%
\bibitem [{\citenamefont {Cao}\ \emph {et~al.}(2022)\citenamefont {Cao},
  \citenamefont {Le~Doussal},\ and\ \citenamefont {Rosso}}]{cao22}%
  \BibitemOpen
  \bibfield  {author} {\bibinfo {author} {\bibfnamefont {X.}~\bibnamefont
  {Cao}}, \bibinfo {author} {\bibfnamefont {P.}~\bibnamefont {Le~Doussal}},\
  and\ \bibinfo {author} {\bibfnamefont {A.}~\bibnamefont {Rosso}},\ }\bibfield
   {title} {\bibinfo {title} {Clusters in an epidemic model with long-range
  dispersal},\ }\href {https://doi.org/10.1103/PhysRevLett.129.108301}
  {\bibfield  {journal} {\bibinfo  {journal} {Phys. Rev. Lett.}\ }\textbf
  {\bibinfo {volume} {129}},\ \bibinfo {pages} {108301} (\bibinfo {year}
  {2022})}\BibitemShut {NoStop}%
\bibitem [{\citenamefont {H.Brezis}\ \emph {et~al.}(2016)\citenamefont
  {H.Brezis}, \citenamefont {L.A.Peletier},\ and\ \citenamefont
  {D.Terman}}]{Brezis}%
  \BibitemOpen
  \bibfield  {author} {\bibinfo {author} {\bibnamefont {H.Brezis}}, \bibinfo
  {author} {\bibnamefont {L.A.Peletier}},\ and\ \bibinfo {author} {\bibnamefont
  {D.Terman}},\ }\bibfield  {title} {\bibinfo {title} {A very singular solution
  of the heat equation with absorption},\ }\href
  {https://doi.org/10.1007/BF00251357} {\bibfield  {journal} {\bibinfo
  {journal} {Arch. Rational Mech. Anal}\ }\textbf {\bibinfo {volume} {95}},\
  \bibinfo {pages} {185} (\bibinfo {year} {2016})}\BibitemShut {NoStop}%
\end{thebibliography}%

\end{document}